\documentclass[reprint,
superscriptaddress,
 amsmath,amssymb,
 aps,
pre,
floatfix,
nofootinbib
]{revtex4-2}

\usepackage{float}
\usepackage{graphicx}% Include figure files
\usepackage{dcolumn}% Align table columns on decimal point
\usepackage{bm}% bold math
\usepackage[colorlinks=true,linkcolor=blue,urlcolor=blue,citecolor=blue]{hyperref}
\usepackage{comment}
\usepackage{footmisc}
\begin{document}

\preprint{APS/123-QED}

\title{ Irreversibility of mesoscopic processes with hydrodynamic interactions }

\author{Biswajit Das}
\email{bd18ip005@iiserkol.ac.in}
\affiliation{Department of Physical Sciences, Indian Institute of Science Education and Research Kolkata, Mohanpur Campus, Mohanpur, West Bengal 741246, India}
\author{Sreekanth K Manikandan}
\email{sreekm@stanford.edu }
\affiliation{Department of Chemistry, Stanford University, Stanford, CA, USA 94305}
\author{Shuvojit Paul}
\thanks{Present address: Kandi Raj College, Kandi, Murshidabad, West Bengal 742161, India}
\affiliation{Department of Physical Sciences, Indian Institute of Science Education and Research Kolkata, Mohanpur Campus, Mohanpur, West Bengal 741246, India}
\author{Avijit Kundu}
\thanks{Present address: Department of Physics, Simon Fraser University, Burnaby, British Columbia, V5A 1S6 Canada}
\affiliation{Department of Physical Sciences, Indian Institute of Science Education and Research Kolkata, Mohanpur Campus, Mohanpur, West Bengal 741246, India}
\author{Supriya Krishnamurthy}
\email{supriya@fysik.su.se}
\affiliation{Department of Physics, Stockholm University, Stockholm, Sweden}
\author{Ayan Banerjee}
\email{ayan@iiserkol.ac.in}
\affiliation{Department of Physical Sciences, Indian Institute of Science Education and Research Kolkata, Mohanpur Campus, Mohanpur, West Bengal 741246, India}

\date{\today}% It is always \today, today,

\begin{abstract}

Optically confined colloidal particles, when placed in close proximity, form a dissipatively coupled system through hydrodynamic interactions. The role of such interactions influencing irreversibility and energy dissipation in out-of-equilibrium systems is often not well deciphered. Here, we demonstrate - through the estimation of the entropy production rate - that the nonequilibrium features of the system with such interactions vary depending on the nature of external driving, and importantly, on the level of coarse-graining. Crucially, we show that coarse-graining reverses the dependence of the measured entropy production rate on the strength of the hydrodynamic interactions. Furthermore, we clarify that such interactions do not violate energy balance at the level of individual trajectories, as was believed earlier. 
Our results highlight a previously unnoticed effect of coarse-graining in nonequilibrium systems, and have implications for the inference of entropy production in experimental contexts.

\end{abstract}

\maketitle

%\section*{}
\textit{Introduction.-}  
Hydrodynamic interactions, which are ubiquitous in fluidic environments ~\cite{doi1988theory,happel2012low}, significantly impact the dynamical properties of mesoscale systems \cite{meiners1999direct,dufresne2000hydrodynamic,cui2004anomalous,polin2006anomalous,leoni2010minimal}. These interactions are crucial for the self-organization of biological materials, such as protein folding \cite{yuan2024impact} and can influence multiple cellular properties and processes, including cell morphology, intracellular processes, cell-cell signalling cascades and reaction kinetics \cite{huber2018hydrodynamics}. Furthermore, hydrodynamic couplings facilitate the synchronization of microscopic oscillators associated with the collective motion of cilia or flagella \cite{kotar2010hydrodynamic,curran2012partial}.

Optically trapped colloidal particles at close separations constitute minimal experimental systems where these interactions emerge naturally, and can be studied in a controllable fashion. Interestingly, the resulting hydrodynamics can also be characterized analytically with high accuracy. Indeed, a number of previous studies looked at optically trapped particles in equilibrium as well as near equilibrium states \cite{meiners1999direct,polin2006anomalous,herrera2013hydrodynamic,paul2017direct,paul2019quantitative} and used them as a means to characterize the emergent interactions. For example, correlation functions of position fluctuations of two closely spaced, hydrodynamically coupled colloidal particles show pronounced time-delayed anticorrelation~\cite{meiners1999direct}, while systems with arrays of multiple hydrodynamically coupled particles trapped in optical potentials can behave as elastic media~\cite{polin2006anomalous,cicuta2010hydrodynamic}. Furthermore, hydrodynamically coupled particles are also known to show synchronisation~\cite{kotar2010hydrodynamic}, and exhibit resonance-like phenomenon~\cite{paul2017direct,paul2019quantitative} under a deterministic driving.   

Colloidal particles at close separations where hydrodynamic effects play a role are also very interesting from the point of view of 
stochastic thermodynamics~\cite{seifert2012stochastic,seifert2019stochastic}. 
Their non-equilibrium characteristics have been earlier quantified in terms
of trajectory energetics \cite{berut2014energy, berut2016stationary,berut2016theoretical, krishnamurthy2022synergistic} or area enclosing rates \cite{thapa2024nonequilibrium}. On the basis of these studies, it was seen that some properties of these systems differ from those of colloidal particles with conservative interactions~\cite{thapa2024nonequilibrium}. In fact, from analysis of the trajectory energetics, it was even argued that the non-conservative nature of the hydrodynamic interactions in these systems leads to violations of energy balance ~\cite{berut2014energy,berut2016stationary, berut2016theoretical}. 

In this paper, we take a look at this class of systems using a single classifier for their out-of-equilibrium behaviour- the total entropy production rate (EPR)~\cite{seifert2005entropy,landi2013entropy,roldan2010estimating,li2019quantifying, roldan2021quantifying}. For our system of interest, we determine the EPR for both the entire system when all degrees of freedom are visible as well as a version of it when one degree of freedom is inaccessible. We find that, in these two cases, the EPR shows diametrically opposite trends as a function of the distance between the particles. In the former case, the EPR increases as the distance between the particles increases (decreasing interaction strength), while in the latter case, it decreases with increasing inter-particle distance. To our knowledge, the fact that coarse-graining can lead to such an effect has not been noted earlier. Furthermore, we clarify how to carry out the trajectory energetics for these systems, which keep track of all the heat flows, obeying the first law at the level of individual trajectories.

\textit{Results.-} We first consider a system with two hydrodynamically coupled particles trapped in two separate parabolic potentials while the mean position of one of the traps is modulated by the active Ornstein-Uhlenbeck (OU) process~\cite{pal2013work,gomez2010steady,manikandan2021quantitative, das2023enhanced}. This particular form of external driving is often used to model the response of an active bath~\cite{maggi2014generalized}. In this system - which we call \textit{`OU-noise driven model'} in further discussions -  detailed balance is always violated because of the driving, leading to a non-equilibrium steady state (NESS) over time, even when the two particles are well separated with minimal hydrodynamic interaction. In addition, the strength of the hydrodynamic interaction can be tuned by changing the mean separation between the two particles. 

To realise the system experimentally, two trapping potentials with a separation $d$ are created by tight-focusing two separate Gaussian beams (wavelength, $\lambda = 1064\ nm$) - emanating from solid-state lasers of perpendicular linear polarisation states - through an objective lens of high numerical aperture (NA $= 1.3$, $100X$, \textit{oil-immersion}) positioned in a conventional inverted microscope (\textit{Olympus XI}). One of the trapping beams is passed through an acousto-optic modulator (AOM) which is externally modulated by active OU noise of different amplitudes and timescales. Polystyrene microparticles (diameter, $2a = 3\ \mu m$) - sparsely dispersed in double-distilled water - are trapped inside a sample holder placed on the stage of the microscope. To detect the position fluctuations of the trapped microparticles, two detection beams of different wavelengths ( $650\ nm$ and $785\ nm$),  co-propagating with the trapping beams, are loosely focused onto the trapped particles. The back-scattered light from both particles is separately projected on separate `balanced-detection' systems constructed using high-gain photo-diodes \cite{bera2017fast}. In this way, the trajectories of both particles are independently recorded at $10\ kHz$ for $50s$ and used for further analysis. 
\begin{figure}
    \centering
    \includegraphics[width=0.5\textwidth]{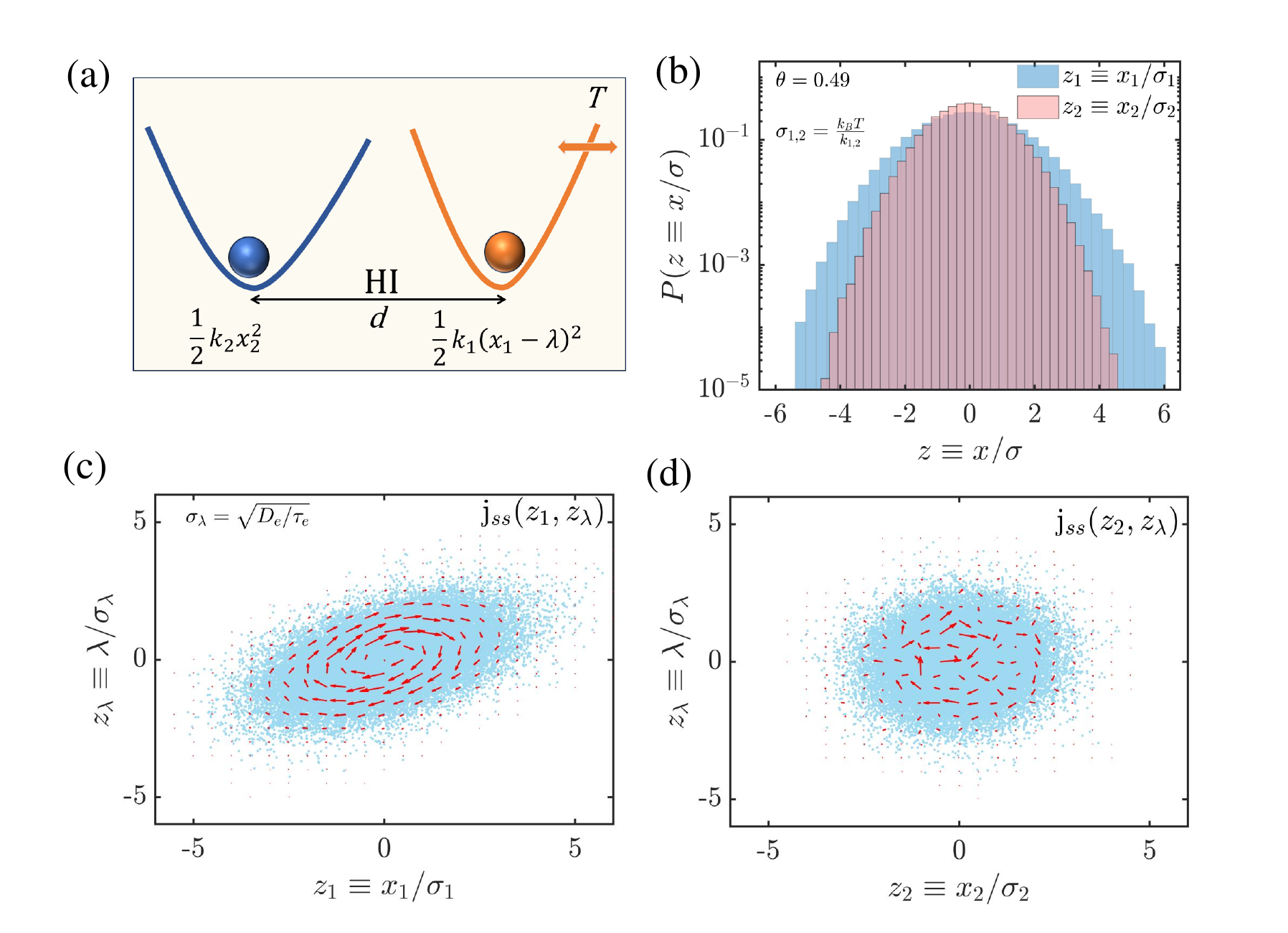}
    \caption{(a) Schematic of the hydrodynamically coupled particles in dual traps of different stiffness constants.  The mean position of the trap with stiffness constant `$k_1$' is modulated with an Ornstein-Uhlenbeck noise (`$\lambda(t)$') while the trap with stiffness constant `$k_2$' remains unperturbed. $k_1 = 17.2\pm0.2\ pN/\mu m $ and $k_2 = 12\pm0.4\ pN/\mu m$ are fixed throughout the experiments. (b) The probability density functions of normalised positional fluctuations of the particles trapped in two traps ($d = 4.2\ \mu m$) are plotted. The experimentally recorded fluctuations of both particles are normalised with the equilibrium standard deviations ($\sigma_i = \sqrt{k_B T/k_i}$) of the corresponding traps. (c) The non-zero probability current corresponding to the particle in the driven trap is shown. (d) The probability current corresponding to the particle in the fixed trap appears to be non-zero as well.}
    \label{fig:1}
\end{figure}

The dynamics of the system can be written as a multidimensional linear \textit{Langevin} equation: $\mathbf{\dot{x}} (t) = - \mathbf{F}\cdot\mathbf{x}(t) + \boldsymbol{\xi}(t)$, with $\langle \boldsymbol{\xi}(t):\boldsymbol{\xi}(t^\prime)\rangle = 2\delta(t-t^\prime)\mathbf{D}$. Here, $\mathbf{x}(t) \equiv [x_1(t), x_2(t), \lambda(t)]^T$ consists of the fluctuating positions ($x_1(t), x_2(t)$) of the two particles measured with respect to the center of each optical trap having stiffness constants $k_1$ and $k_2$, and $\lambda(t)$ is the OU noise that is exponentially correlated with the relaxation timescale $\tau_e$ ($>\gamma/k_1$) and amplitude $D_e$ with $\langle \lambda(t) \lambda(t^\prime) \rangle = \frac{D_e}{\tau_e}\exp (-\frac{t-t^\prime}{\tau_e})$. The vector $\boldsymbol{\xi}(t) \equiv [\eta_1(t), \eta_2(t), \eta_3(t)]^T$ contains the random \textit{Brownian} forces with $\langle \eta_i (t) \rangle = 0$ and $\langle \eta_i (t) \eta_j(s)\rangle = \delta_{ij} \delta (t-s)$ for any two times $t$ and $s$. 
The drift ($\mathbf{F}$) and diffusion ($\mathbf{D}$) tensors can be expressed in terms of the hydrodynamic coupling constant $\epsilon \equiv (3a/2d) - (a/d)^3$~\cite{berut2016stationary} and the Stokes friction coefficient of the medium $\gamma \equiv 6\pi\eta a$ with viscosity $\eta~(=8.9\times10^{-4}~$Pa.s), such that
\begin{equation}
    \mathbf{F} = 
    \begin{pmatrix}
        k_1/\gamma & \epsilon k_2/\gamma &- k_1/\gamma \\
       \epsilon k_1/\gamma & k_2/\gamma & -\epsilon k_1/\gamma \\
       0 & 0 & 1/\tau_e
       \end{pmatrix}, \
   \mathbf{D} = \begin{pmatrix}
      D_0  &  \epsilon D_0  &0\\
        \epsilon D_0  &  D_0 & 0 \\
       0 & 0 & D_e/\tau_e^2
       \end{pmatrix},
       \label{eq:drift_diffusion_ou}
\end{equation}
as $D_0 = k_B T/\gamma$ ($k_B$ is the Boltzmann's constant). Note that hydrodynamic interactions make the system \textit{non-multipartite} with a non-diagonal diffusion matrix~\cite{leighton2024jensen}.
The rationale behind the dynamical equations is explicitly discussed in SI~\ref{ap1:dynamics}. 

\begin{figure*}[t]
    \centering
\includegraphics[width=0.99\textwidth]{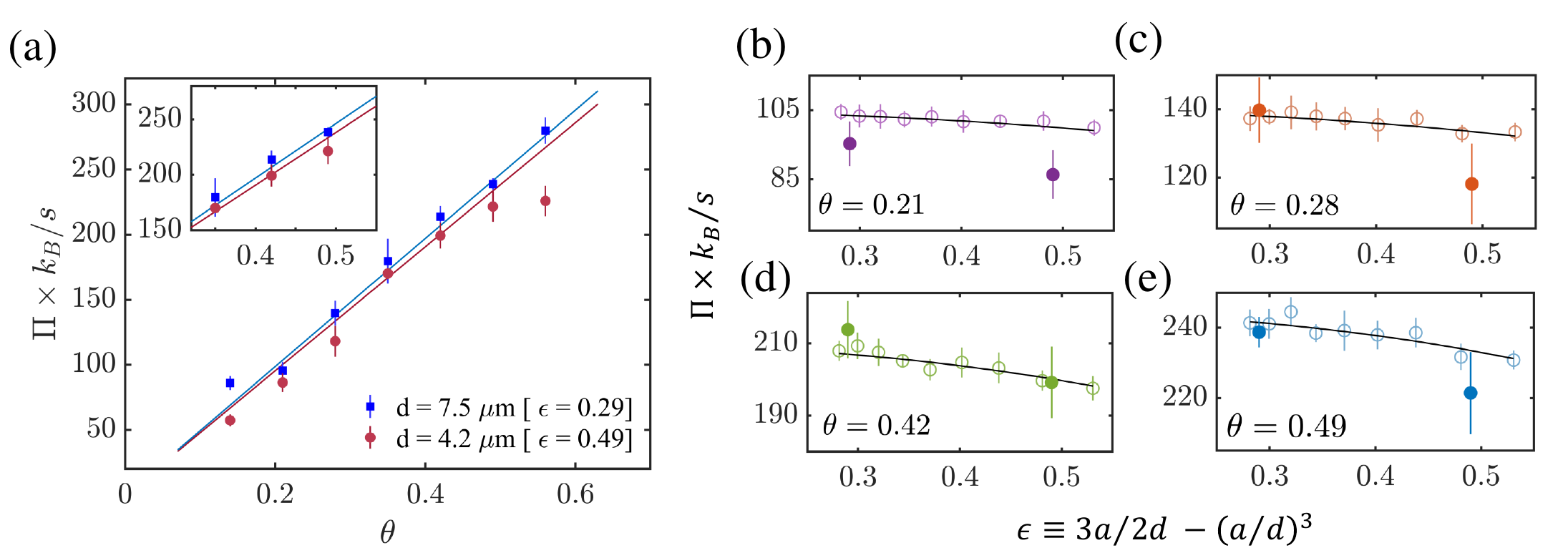}
    \caption{\textit{Entropy production rate ($\Pi$) of the `OU-noise driven model'}: (a) Inferred $\Pi$ from the experimental data monotonically increases with the strength of the external noise (parametrised as $\theta$) at two different separations between the particles. (b) -(e) The EPR of the system decreases with the increasing hydrodynamic interaction ($\epsilon$) as shown for different noise strengths. `\textit{Filled}' and `\textit{open}' symbols are estimated from experimentally and numerically obtained trajectories, respectively. Corresponding theoretical estimates (Eq.~\ref{eq:epr_ou}) are shown with solid lines. The data points are obtained as the mean of the entropy production rates estimated from multiple experimental (or numerical) trajectories and the corresponding standard deviations are shown as error bars.  Details concerning the numerical simulations are provided in SI~\ref{ap5:numerical}.}   
     \label{fig:2}
\end{figure*}

A schematic of the system of our interest is shown in Fig.\ref{fig:1}(a). We primarily focus on the entropy production rate of the system after it reaches a nonequilibrium steady state, defined by the joint probability distribution ($P_{ss}(\mathbf{x})$) and non-zero probability current ($\mathbf{j}_{ss}(\mathbf{x})$). The marginal steady-state distributions of both the particles in the `driven' ($x_1$) and `fixed' ($x_2$) traps are Gaussian as shown in Fig.\ref{fig:1}(b). The variance of the particle in the `driven' trap is expectedly larger than the same in a fixed trap for a typical experimental trajectory. The recorded positional fluctuations of both particles are expressed as dimensionless quantities $z_{1,2} \equiv x_{1,2}/\sigma_{1,2}$ as $\sigma_{1,2} = \sqrt{k_B T/k_{1,2}}$. The non-zero probability current (as shown in Fig.\ref{fig:1}(c)) is strongly prevalent in $(x_1,\lambda)$ space as the particle in the first trap is driven by the OU noise. However, the probability current also prevails in the $(x_2,\lambda)$ space (Fig.\ref{fig:1}(d)), even though the external noise does not directly drive the particle in the second trap. This is indicative of the appearance of induced nonequilibrium dynamics through hydrodynamic interactions, as also observed in \cite{thapa2024nonequilibrium}.

To quantify the effects of these interactions, we perform the experiments at two different separations ($d = 4.2\ \mu m$ and $d = 7.5\ \mu m$) between the traps ($k_1 = 17.2 \pm 0.2\ pN/\mu m$ and $k_2 = 12 \pm 0.4\ pN/\mu m$)  by varying the strength of the external noise (parameterized by $\theta = \frac{D_e/\tau_e^2}{D_0}$) while fixing $\tau_e = 4~ms$  throughout the experiments.  We then estimate the EPR from the experimental and numerical data using the short-time inference technique \cite{manikandan2020inferring,van2020entropy,otsubo2020estimating,manikandan2021quantitative,das2022inferring} based on the thermodynamic uncertainty relation \cite{barato2015thermodynamic}. Using this method, the entropy production rate can be calculated as,
\begin{equation}
    \Pi = \lim_{dt \rightarrow 0} \max_{J} \left[\frac{2 k_B \langle J \rangle^2}{dt \ \text{Var}(J)}\right],
    \label{eq:TUR_J}
\end{equation}
where $J$ is a weighted scalar current - that can be computed from the time-discretised experimental or numerical trajectory data ($\mathbf{x}$) sampled at an interval of $dt$.
This technique is model-independent, and particularly advantageous for estimating the EPR from experimental trajectories, as it does not require any calibration factor to transform experimental measurements into positional units \cite{manikandan2021quantitative}. The technique is also briefly discussed in SI~\ref{ap2:inference_tech}.

We find that the EPR increases monotonically with $\theta$ at both separations (Fig.\ref{fig:2}(a)). However, the entropy generation rate for the separation of $d = 4.2\ \mu m$ at any $\theta$ is found to be slightly lower than the same for $d = 7.5\ \mu m$ as shown in Figs.~\ref{fig:2}(b) - \ref{fig:2}(e). This observation suggests that if the separation between two particles is reduced, the total EPR of the system will also decrease, even though the hydrodynamic coupling strength is enhanced in this process.
  
The analytically calculated total entropy production rate (in units of $k_B /s$) for the system ~\cite{seifert2005entropy},   
\begin{equation}
\begin{split}
    \Pi =& \int d\mathbf{x} \frac{\mathbf{D}^{-1} \mathbf{j}_{ss}^2(\mathbf{x})}{P_{ss}(\mathbf{x})}\\
    =&  \frac{D_e k_1^2 (\gamma +(1-\epsilon^2) k_2 \tau_e)}{D_0 \gamma \tau_e(\gamma^2 + \gamma(k_1 +k_2)\tau_e + (1-\epsilon^2)k_1k_2\tau_e^2)},
\end{split}
\label{eq:epr_ou}
\end{equation}
also corroborates our observation as shown in the plots of Fig.\ref{fig:2}. If the hydrodynamic interaction becomes negligible ($\epsilon \rightarrow 0$), the entropy production rate becomes,
\begin{equation}
    \label{eq:epr_ou_ep0}
     [ \Pi]_{\epsilon \rightarrow 0} = \frac{D_e k_1^2}{D_0 \gamma \tau_e(\gamma + k_1 \tau_e) },
\end{equation}
which is greater than $\Pi$ - indicating the reduction of entropy production rate in the presence of hydrodynamic interactions. 
This occurs because the motion of the `driven' particle gets constricted due to the proximity of the other particle; this results in a lower entropy production rate.
Similar observations were made in Ref.~\cite{manikandan2021quantitative}, where the energy dissipation in a driven particle was found to be reduced due to hydrodynamic flows close to a microbubble. 

Colloidal entities and other nanoscale probes have often been used recently, as a means to characterise nonequilibrium characteristics of complex systems~\cite{basu2015statistical,bacanu2023inferring,seyforth2022nonequilibrium,di2024variance}.
In that context, note that we could also think of our two-particle system as a composite system consisting of the driven particle ($x_1$) - which has an intrinsic energy dissipation rate $[\Pi]_{\epsilon \rightarrow 0}$ -  being the system of interest, and the other particle ($x_2$) being a probe brought close to it. Our findings so far demonstrate that
hydrodynamic interactions with a probe, which by itself does not dissipate energy, could significantly reduce the energy dissipation of the nonequilibrium system of interest. As a result, one has to take such considerations into account when estimating the EPR of a complex system with the help of a probe.

\begin{figure}[t]
    \centering
    \includegraphics[width = 0.49\textwidth]{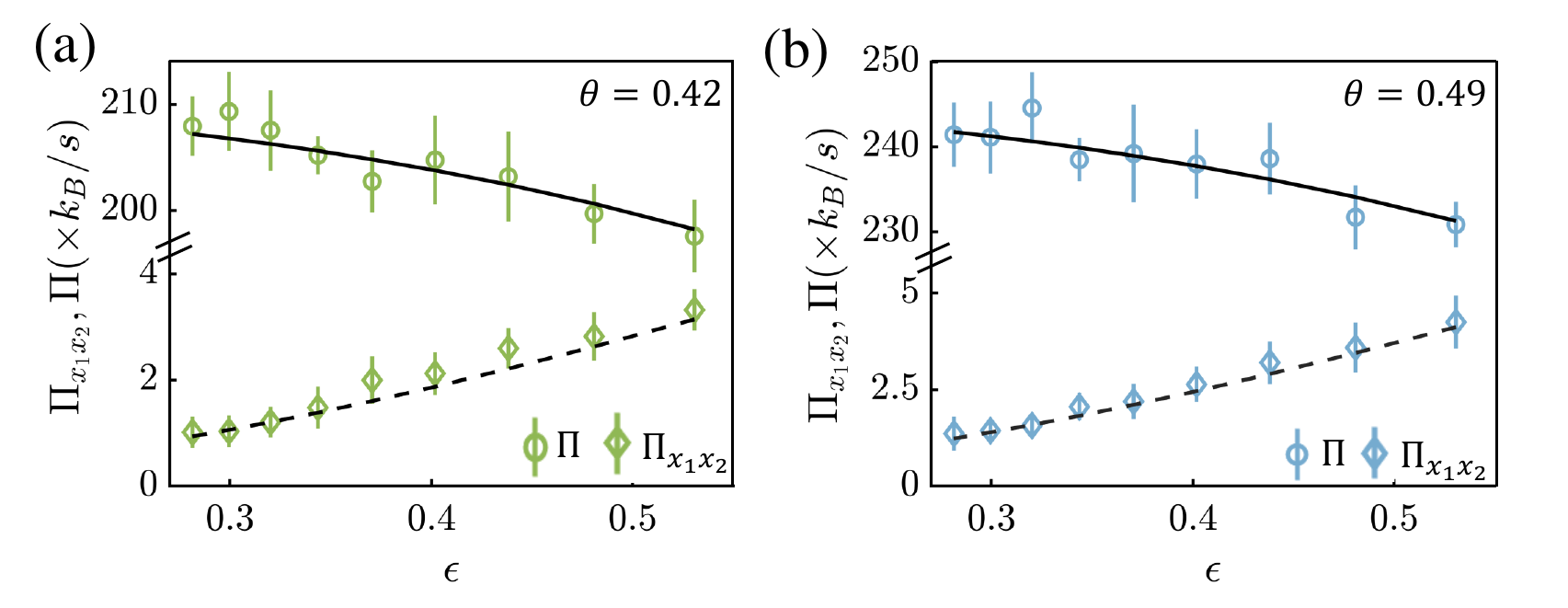}
    \caption{(a)-(b): Partial entropy production rates ($\Pi_{x_1x_2}$) estimated from numerical trajectories are plotted as a function of the hydrodynamic interaction strength($\epsilon$). $\Pi_{x_1x_2}$ is significantly lower and shows the opposite dependence on $\epsilon$ compared to the total entropy production rate of the system. Black solid lines represent the theoretical estimation of the total entropy production rate of the system (Eq.~\eqref{eq:epr_ou}) and the dashed lines indicate the theoretical estimation of the partial entropy production rate (Eq.~\eqref{ap:eq_reduced_epr} of SI~\ref{ap3:partial_ou}).}
    \label{fig:3n}
\end{figure}

We now focus on another interesting aspect that relates to estimations of the EPR when not all the degrees of freedom are accessible~\cite{mehl2012role,dieball2024perspective}. Towards this, consider that the OU noise ($\lambda(t)$) in our system is not accessible, and we are only able to track the positions of the two colloidal particles $x_1 (t)$ and $x_2 (t)$.
Following the technique introduced in Ref.~\cite{nicoletti2024tuning, mura2018nonequilibrium}, the partial entropy production rate ($\Pi_{x_1x_2}$) of our system with only $x_1(t)$ and $x_2(t)$ as available degrees of freedom can be obtained analytically by substituting the diffusion matrix $\mathbf{D}$ and the steady state covariance matrix $\mathbf{C}$ by their respective submatrices   $\mathbf{D}_{x_1 x_2}$ and $\mathbf{C}_{x_1 x_2}$. In addition, the drift matrix $\mathbf{F}$ needs to be replaced by the reduced drift matrix $\mathbf{F}^{red}_{x_1 x_2}$ as,
\begin{equation}
    \mathbf{F}^{red}_{x_1 x_2} = \mathbf{F}_{x_1 x_2} + \mathbf{B}\mathbf{C}_{x_1 x_2}^{-1}, \\\\\\\\\ (\mathbf{B})_{ij} = (\mathbf{F})_{i\lambda} (\mathbf{C})_{j\lambda}
    \label{eq:reduced_F}
\end{equation}
with $i, j = x_1, x_2$. Details of this calculation are provided in SI~\ref{ap3:partial_ou} (Eq.~\eqref{ap:eq_epr_matrix}-Eq.~\eqref{ap:eq_reduced_epr}. We find that the estimated partial entropy production rate is significantly lower than the total entropy production rate of the system as expected. However, somewhat counter-intuitively, $\Pi_{x_1x_2}$ increases with the increase in strength of the hydrodynamic interactions as shown in Figs.~\ref{fig:3n}(a)-\ref{fig:3n}(b), hence displaying the opposite dependence on distance (interaction strength, $\epsilon$) compared to the total entropy production rate of the system. Note that another measure of the nonequilibrium nature in the $x_1-x_2$ phase space is the area enclosing rate \cite{mura2018nonequilibrium}. Interestingly, in Ref.~\cite{thapa2024nonequilibrium}, this quantifier was further studied for a theoretical model consisting of three hydrodynamically interacting particles, where it was observed that the area enclosing rate decreased as a function of the distance for any two-dimensional projection of the system. However, no comparison was made with the EPR of the full three-dimensional system.

Now, for the model studied so far, we have found that the total EPR ($\Pi$) increases as $\epsilon \rightarrow 0$ or equivalently as the distance between the particles $d$ increases. This could be understood from the fact that the presence of the non-driven particle was a constraining presence on the driven one, so that the farther the particles were away from each other, the larger was the EPR of the system. 
The obvious question that arises now is - how then can we understand that when $\lambda(t)$ becomes a hidden degree of freedom, the entropy of the reduced system $\Pi_{x_1,x_2} $   actually decreases as the particles move further apart?
We argue below on heuristic grounds that this feature may be interpreted through the emergence of unequal effective temperatures in the coarse-grained state space (marginalised over the additional stochastic noise $\lambda(t)$).

Consider the following model (which we henceforth refer to as the `\textit{two-temperature model}' )
with drift ($\mathbf{F}_{tw}$) and diffusion ($\mathbf{D}_{tw}$) tensors of following forms \cite{berut2014energy} -
\begin{equation}
    \mathbf{F}_{tw} = 
    \begin{pmatrix}
        k_1/\gamma & \epsilon k_2/\gamma\\
       \epsilon k_1/\gamma & k_2/\gamma  \\
       \end{pmatrix},\
   \mathbf{D}_{tw} = \begin{pmatrix}
      \frac{k_B (T + \Delta T)}{\gamma} &  \epsilon \frac{k_B (T + \Delta T)}{\gamma}\\
     \epsilon \frac{k_B (T + \Delta T)}{\gamma}  &   \frac{k_B (T +  \epsilon^2 \Delta T)}{\gamma}\\
       \end{pmatrix},
       \label{eq:drift_diffusion_tw}
\end{equation}
where the particle in the trap of stiffness $k_1$ feels the temperature $T+\Delta T$ and the other particle in the trap of stiffness $k_2$ feels a lower temperature $T$. The underlying rationale for the dynamical equations of this system is elegantly explained in Refs.~\cite{berut2014energy,yolcu2016linear}. This model is also experimentally viable and has been studied in detail in several Refs.~\cite{berut2014energy,berut2016stationary}. The total entropy production rate (in units of $k_B/s$) of this system (corresponding to $x_1-x_2$ phase space) can be analytically computed:
\begin{equation}
    \label{eq:epr_w}
    \Pi_{tw} = \frac{k_2^2 \Delta T^2 \epsilon^2 (1-\epsilon^2)}{(k_1 + k_2) \gamma T(T + \Delta T) }.
\end{equation}
$\Pi_{tw}$ is found to be non-monotonic in $\epsilon$ with a maxima at $\epsilon = 1/\sqrt{2} \equiv 0.707 $. This indicates that the entropy production rate of the system is enhanced with increased hydrodynamic coupling up to a certain separation beyond which it goes down even though the interaction is enhanced. Note however that the interparticle separation in this latter phase is extremely low as compared to the particle size ($a \sim 1.5~ \mu m$ ) and further higher-order corrections to the hydrodynamic interactions are expected to play a role. Hence the physically accessible regime is the one where the EPR decreases as a function of increasing interparticle distance (decreasing $\epsilon$) as in the coarse-grained version of the \textit{OU-noise driven model} in two dimensions.
\begin{figure*}
    \centering
    \includegraphics[width = 0.99\textwidth]{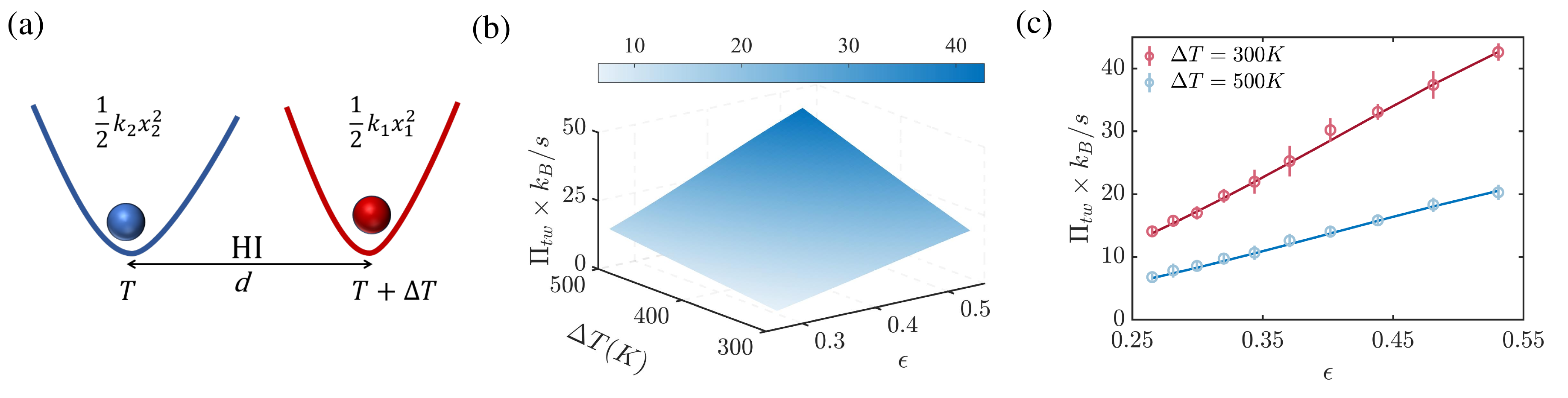}
    \caption{\textit{`Two-temperature model'}: (a) Schematic of the hydrodynamically coupled particles in dual traps of different stiffness constants.  The particle trapped in the potential with stiffness constant `$k_1$' is at temperature $T+\Delta T$ while the particle in the trap with stiffness constant `$k_2$' remains at temperature $T$. (b) Theoretical estimation (Eq.~\ref{eq:epr_w}) of the entropy production rate of the system is plotted as a function of $\Delta T$ and hydrodynamic coupling strength $\epsilon$. [$k_1 = 17.2 pN/\mu m$, $k_2 = 12 pN/\mu m$] (c)  Estimation of entropy production rate as a function of $\epsilon$ -  using the short-time inference scheme from numerical trajectories of the system for fixed $\Delta T$. The solid lines indicate corresponding theoretical predictions (Eq.~\ref{eq:epr_w}). The data points are obtained as the mean of the entropy production rates estimated from multiple numerical trajectories and the corresponding standard deviations are shown as error bars. Details of numerical simulation are provided in SI~\ref{ap5:numerical}.}  
    \label{fig:3}
\end{figure*}
The analytical and numerical estimation of the total entropy production rate for this system as a function of $\Delta T$ and $\epsilon$ are shown in Fig.~\ref{fig:3}(b) and in Fig.~\ref{fig:3}(c) for interparticle separations ($d$) varying from $3.8~\mu m$ - $8.3~ \mu m$. 

It is clear that the EPR in the \textit{two-temperature model} qualitatively exhibits the same features as
$\Pi_{x_1,x_2} $, motivating our analogy between the two cases. The effect of the unseen driven degree of freedom  $\lambda(t)$ emerges as an effective temperature difference between the two particles. However, the fact that this reverses the dependence of the EPR on the inter-particle distance is extremely interesting, since it implies that such trends which may be found experimentally from tracking a coarse-grained degree of freedom do not necessarily imply the same trend when all degrees of freedom are measured.

\textit{Discussions on trajectory energetics:} 
We now turn our attention to another important aspect of our system - namely the trajectory energetics, which can be estimated from the time evolution of the degrees of freedom of the system. 
Now, since the noises for the different degrees of freedom are cross-correlated in the presence of hydrodynamic interactions,  the system becomes non-multipartite, and it is not completely obvious how to obtain quantities such as heat or work unambiguously from the trajectories~\cite{leighton2024jensen}.
Indeed, previous studies have suggested that energy balance itself does not hold at the level of individual trajectories or at the level of ensemble averages, for systems having hydrodynamic interactions~\cite{berut2014energy,berut2016stationary,berut2016theoretical}. However, we argue that this claim is incorrect, and that the degrees of freedom of the system account for all the energy that is dissipated. 
 For the \textit{OU-noise driven system}, using standard definitions \cite{seifert2012stochastic}, we can identify the work done ($W_\tau$) in an individual realization of duration $\tau$ as, $W_{\tau} =  \int_0^\tau dt\; \partial_\lambda  V(x_1,x_2,\lambda )\circ \dot{\lambda}
        = k_1\int_0^\tau dt\;  (\lambda - x_1)\circ\dot{\lambda}$. Here, $V(x_1,x_2,\lambda) = 1/2(k_1(x_1 - \lambda)^2 + k_2 x_2^2)$ and $\circ$ denotes a \textit{Stratanovich} product. Similarly, we can identify heat dissipated along a trajectory as $Q_{\tau} =  \int_0^\tau dt\; \partial_{\bm x}  V(x_1,x_2,\lambda ) \circ \dot{\bm x}$ with $\bm x = (x_1,x_2)$. 
This definition leads to a decomposition of heat for individual particles as, 
        $Q_{\tau} =  \int_0^\tau dt\; \partial_{\bm x}  V(x_1,x_2,\lambda ) \circ \dot{\bm x} = \int_0^\tau dt\; \partial_{ x_1}  V(x_1,x_2,\lambda ) \circ \dot{ x}_1 +  \int_0^\tau dt\; \partial_{ x_2}  V(x_1,x_2,\lambda ) \circ \dot{ x}_2 \equiv Q_{1 \tau} +Q_{2 \tau}$.
        Their sum gives, $W_{\tau} +Q_{\tau} = W_{\tau} +Q_{1\tau} +Q_{2\tau} = V(x_1(\tau),x_2(\tau),\lambda(\tau)) - V(x_1(0),x_2(0),\lambda(0))$, which establishes the energy balance for the hydrodynamically coupled two-particle system. Naturally, energy balance then also holds at the level of ensemble averages. Furthermore, as shown in SI~\ref{ap4:energetics} (Eq.~\eqref{eq:mediumep_path}-Eq.~\eqref{eq:mediumep_path_f}), we can also verify the relation $ \langle\dot{Q}\rangle = D_0 \gamma \langle \Delta \dot{S}_{m}\rangle/k_B$ where $\langle \Delta \dot{S}_{m}\rangle$ is the medium entropy production rate. 
        However, if we follow instead the prescription in
        Refs.\ \cite{berut2014energy, berut2016stationary, berut2016theoretical} for identifying heat flows for individual particles, then we encounter some inconsistencies. In SI~\ref{ap4:energetics}, we show that applying the prescription in Refs.\ \cite{berut2014energy, berut2016stationary, berut2016theoretical} to the \textit{OU noise driven model} leads to \textit{a)} the wrong conclusion that energy is not conserved in this simple system and \textit{b)} a discrepancy between the medium entropy estimates computed in two different ways. Hence, we conclude that the prescription for identifying heat in Refs.\ \cite{berut2014energy, berut2016stationary, berut2016theoretical}, originally applied to the \textit{two-temperature model}, is incorrect. We also conclude that hydrodynamic interactions alone do not cause a violation of energy balance. 
        
         Now, to correctly identify $Q_1$ and $Q_2$ at the level of individual trajectories for the \textit{two-temperature model} (there is no explicit external driving, so $W_\tau = 0$), we adopt an inverse approach, leveraging the standard expression for medium entropy production ~\cite{seifert2005entropy},
        \begin{align}
            \begin{split}
                \Delta S_{m(\tau)}^{tw} &\equiv k_B \int_0^\tau dt\ \mathbf{D}_{tw}^{-1} (\mathbf{F}_{tw}\mathbf{x}) \circ \dot{\mathbf{x}}\\ =& \int_0^\tau dt\ \frac{ (k_1 T x_1-\Delta T k_2 x_2 \epsilon )\circ \dot{x}_1+k_2 (\Delta T+T)x_2 \circ \dot{x}_2}{T (\Delta T+T)}, 
            \end{split}
        \end{align}
        and its relation to the heat dissipated to each reservoir: $\Delta S_{m(\tau)}^{tw} = -\frac{Q_{1\tau}}{T} - \frac{Q_{2\tau}}{T + \Delta T}$. 
        This, along with the requirement of energy conservation, i.e. $Q_{1\tau}  + Q_{2\tau} =  (V(x_1(\tau),x_2(\tau)) - V(x_1(0),x_2(0))$, where, $V(x_1,x_2) = 1/2 (k_1 x_1^2 + k_2 x_2^2)$, gives us two independent equations for $Q_{1\tau}$ and $Q_{2\tau}$ that can be solved straightforwardly (see SI~\ref{ap4:energetics} for the relevant expressions given in Eq.~\eqref{eq:SQ1}-Eq.~\eqref{eq:SQ2}). It remains an interesting open problem, though, to understand how to identify heat flows in the general case with a non-diagonal diffusion matrix and several reservoirs at different temperatures.
 
\textit{Conclusions.-} In summary, we have demonstrated the role of hydrodynamic interactions in shaping the energetics and irreversibility of mesoscopic processes, and elucidated the effects of coarse-graining the nonequilibrium drive. Crucially, we have shown that coarse-graining can completely reverse the trend of the measured entropy production rate with the strength of interactions. To the best of our knowledge, this has not been noticed earlier.
Our approach is based on the quantification of the total entropy production rate from experimental data, and is scalable to systems with many degrees of freedom. 

The results presented here can be straightforwardly applied to polygonal arrays of particles~\cite{cicuta2010hydrodynamic}, or colloidal crystals to design novel non-equilibrium materials~\cite{polin2006anomalous}, and to correlate the emergent properties to the underlying non-equilibrium state~\cite{caprini2023entropy}. It will also be interesting to see if these interactions can be harnessed to design artificial motors at the nanoscale.  The impact of such interactions on information transfer in biochemical signalling networks \cite{nicoletti2021mutual,hahn2023dynamical}, as well as on optimising protocols for non-equilibrium control problems \cite{aurell2011optimal,chennakesavalu2023unified, chennakesavalu2024adaptive,rose2024role} will also be very interesting to investigate.

\textit{Acknowledgements.-}The work is supported by IISER Kolkata, and the Science and Engineering Research Board (SERB), Department of Science and Technology, Government of India, through the research grant CRG/2022/002417. BD is thankful to the Ministry Of Education of
Government of India for financial support through the Prime Minister’s Research Fellowship (PMRF) grant. SKM acknowledges the Knut and Alice Wallenberg Foundation for financial support through Grant No. KAW 2021.0328. S.K acknowledges the support of the Swedish Research  Council through the grant 2021-05070.

%\bibliography{references.bib}
%\nocite{}


\begin{widetext}
\section*{Supplementary Materials}
\setcounter{equation}{0}
\makeatletter
\renewcommand\thetable{S\@arabic\c@table}
\renewcommand \theequation{S\@arabic\c@equation}
\makeatother

\subsection{OU-noise driven model: Dynamical descriptions}
\label{ap1:dynamics}
The dynamics of the two identical microscopic particles (radius $a$) trapped in parabolic potentials separated by a distance $d$ with different stiffness constants $k_1$ and $k_2$ can be expressed by coupled $Langevin$ equations \cite{meiners1999direct,berut2016stationary,paul2017direct},
\begin{equation}
\label{eq:ap_dyn_eq1}
    \begin{split}
    \dot{x}_1(t) =& \mathcal{H}_{11}[-k_1 x_1(t) + \xi_1(t)] + \mathcal{H}_{12}[-k_2 x_2(t) + \xi_2(t)]\\
     \dot{x}_2(t) =& \mathcal{H}_{21}[-k_1 x_1(t) + \xi_1(t)] + \mathcal{H}_{22}[-k_2 x_2(t) + \xi_2(t)],
    \end{split}
\end{equation}
where $\mathcal{H}_{ij}$ $(i,j = 1,2)$ are the constant elements of a hydrodynamic coupling tensor of the form $\mathbf{\mathcal{H}} = \begin{pmatrix}
1/\gamma & \epsilon/\gamma\\
\epsilon/\gamma & 1/\gamma\\
\end{pmatrix}$. Here $x_1(t)$ and $x_2(t)$ are displacements from the minimum of the individual traps. It is further assumed that the displacements of the particles ($x_1(t)$, $x_2(t)$) are small compared to the mean separation between the traps ($d$), and the parameter $\epsilon \equiv \frac{3 a}{2 d} - (\frac{a}{d})^3$ denotes the hydrodynamic coupling coefficient taken \cite{meiners1999direct,herrera2013hydrodynamic,berut2016stationary} for longitudinal motions. Here $\gamma \equiv 6\pi\eta a$ is the viscous drag coefficient of the medium. The terms $\xi_1(t)$ and $\xi_2(t)$ are delta-correlated random Brownian forces such that $\langle \xi_i(t)\rangle =0,\ \langle \xi_i(t)\xi_j(t)\rangle = 2 k_B T (\mathcal{H})^{-1}_{ij} \delta(t-t^\prime)$. $k_B$ is the Boltzmann's constant.

Without any external perturbation, the coupled particles will be in thermal equilibrium with the bath of temperature $T$. Following  Ref. \cite{berut2014energy}, Eq.\eqref{eq:ap_dyn_eq1} can be rewritten as,
\begin{equation}
\label{eq:ap_dyn_eq2}
    \begin{split}
        \dot{x}_1(t)&= b_1 (x_1(t), x_2(t)) + \eta_1(t) \\
      \dot{x}_2(t)&= b_2 (x_1(t), x_2(t)) + \eta_2(t), 
    \end{split}
\end{equation}
with 
\begin{equation}
\label{eq:ap_dyn_eq3}
    \begin{split}
      b_i(x_i,x_j) &= -\frac{1}{\gamma} k_i x_i - \frac{\epsilon}{\gamma} k_j x_j,\\
      \eta_1 &= \frac{1}{\gamma}(\xi_1 + \epsilon \xi_2),\\
      \eta_2 &= \frac{1}{\gamma}(\xi_2 + \epsilon \xi_1).
    \end{split}
\end{equation}
The Fokker-Planck equation corresponding to the time evolution of the joint probability distribution ($\rho_{eq}(x_1,x_2,t)$) of this (thermally) equilibrium configuration can be written as,
\begin{equation}
    \label{eq:ap_dyn_fp1}
    \partial_t\rho_{eq}(x_1,x_2,t) = -\partial_{x_1}(b_1 \rho_{eq}) - \partial_{x_2}(b_2 \rho_{eq}) + \alpha_{11}\partial_{x_1}^2\rho_{eq}+ \alpha_{22}\partial_{x_2}^2\rho_{eq} + 2 \alpha_{12}\partial_{x_1 x_2}^2\rho_{eq} .
\end{equation}
$\alpha_{11} = \alpha_{22} \equiv \frac{k_B T}{\gamma} $ and $\alpha_{12} = \alpha_{21} \equiv \epsilon \frac{k_B T}{\gamma} $ are the elements of the equilibrium diffusion matrix ($\mathbf{D}_{eq}$) of the system such that $\langle \eta_i(t) \eta_j(t^\prime)\rangle = 2 (\mathbf{D}_{eq})_{ij} \delta(t-t^\prime)$.\\

To drive the system out of equilibrium, the particle trapped in the potential with stiffness constant $k_1$ is modulated with an  Ornstein-Uhlenbeck (OU) noise ($\lambda(t)$). The external modulation is exponentially correlated with the relaxation timescale $\tau_e$ and amplitude $D_e$ as $\langle \lambda(t) \lambda(t^\prime) \rangle = \frac{D_e}{\tau_e}\exp (-\frac{t-t^\prime}{\tau_e})$  and it is derivable from following dynamical equation, 
\begin{equation}
\label{eq:ap_ou_noise}
    \dot{\lambda}(t) = -\frac{\lambda(t)}{\tau_e} + \frac{\sqrt{2D_e}}{\tau_e}\eta_3(t), ~~~~~ \langle \eta_3(t)\rangle = 0, \langle \eta_3(t)\eta_3(t^\prime)\rangle = \delta(t-t').
\end{equation} 
Now, the dynamics of the perturbed system with the external OU modulation can be expressed with a system of \textit{Langevin} equations with $\lambda(t)$ as another degree of freedom in addition to $x_1(t)$ and $x_2(t)$ such that,
\begin{equation}
\label{eq:ap_dyn_neq1}
    \begin{split}
    \dot{x}_1(t) =& \mathcal{H}_{11}[-k_1 (x_1(t) - \lambda (t)) + \xi_1(t)] + \mathcal{H}_{12}[-k_2 x_2(t) + \xi_2(t)]\\
     \dot{x}_2(t) =& \mathcal{H}_{21}[-k_1 (x_1(t) - \lambda (t)) + \xi_1(t)] + \mathcal{H}_{22}[-k_2 x_2(t) + \xi_2(t)]\\
     \dot{\lambda}(t) =& -\frac{\lambda(t)}{\tau_e} + \frac{\sqrt{2D_e}}{\tau_e}\eta_3(t).
    \end{split}
\end{equation}
Here, $\langle\xi_1(t)\eta_3(t')\rangle = \langle\xi_2(t)\eta_3(t')\rangle =0 \implies \langle\eta_1(t)\eta_3(t')\rangle = \langle\eta_2(t)\eta_3(t')\rangle = 0 $.
Moreover, Eq.\eqref{eq:ap_dyn_neq1} can be rewritten in the following matrix form,
\begin{equation}
\label{eq:model_matrix}
    \begin{pmatrix}
        \dot{x}_1(t) \\ \dot{x}_2(t)\\ \dot{\lambda}(t)
    \end{pmatrix}
    = -  \begin{pmatrix}
        k_1/\gamma & \epsilon k_2/\gamma & -k_1/\gamma \\
      \epsilon k_1/\gamma & k_2/\gamma & -\epsilon k_1/\gamma \\
       0 & 0 & \frac{1}{\tau_e}
        \end{pmatrix}
        \begin{pmatrix}
            x_1(t) \\ x_2(t) \\ \lambda(t)
        \end{pmatrix}
        +
       \begin{pmatrix}
            \eta_1(t) \\ \eta_2(t) \\ \eta_3(t)
        \end{pmatrix}
\end{equation}
\begin{equation}
\label{ap:eq_langevin}
            \implies \mathbf{\dot{x}(t)} = -\mathbf{F}.\mathbf{x(t)} + \boldsymbol{\xi(t)},
\end{equation}
 with $\langle \boldsymbol{\xi}(t):\boldsymbol{\xi}(t^\prime)\rangle = 2\delta(t-t^\prime)\mathbf{D}$. Therefore, the drift ($\mathbf{F}$) and diffusion ($\mathbf{D}$) matrices for the \textit{`OU-noise driven model'} are of following forms,
\begin{equation}
\label{eq:ap_eq_neq_FD}
    \mathbf{F} = 
    \begin{pmatrix}
        k_1/\gamma & \epsilon k_2/\gamma &- k_1/\gamma \\
       \epsilon k_1/\gamma & k_2/\gamma & -\epsilon k_1/\gamma \\
       0 & 0 & \frac{1}{\tau_e}
       \end{pmatrix}, \
   \mathbf{D} = \begin{pmatrix}
      D_0  &  \epsilon D_0  &0\\
        \epsilon D_0  &  D_0 & 0 \\
       0 & 0 & \frac{D_e}{\tau_e^2}
       \end{pmatrix},
\end{equation}
with $D_0 = \frac{k_B T}{\gamma}$. It is interesting to note that, the thermal noises of the bath are cross-correlated by the hydrodynamic interaction that leads to the non-diagonal form of the diffusion matrix. This peculiar feature of hydrodynamic interaction also makes the system \textit{non-multipartite}~\cite{leighton2024jensen}. \\
%{\bf In addition, the drift matrix is also asymmetric due to the nonreciprocal nature of the hydrodynamic forces.}

\textit{Steady state distribution}.-
The probability of finding the particle in a configuration $\mathbf{x}$ at a certain time $t$ can be determined in terms of the probability distribution function $P(\mathbf{x},t)$ which follows the Fokker-plank equation of the form
\begin{equation}
    \begin{split}
    \partial_t P(\mathbf{x},t) &= - \nabla \cdot (-\mathbf{F}\mathbf{x}P({\mathbf{x},t}) - \mathbf{D} \nabla P(\mathbf{x},t)) \\
    & \equiv -\nabla \cdot \mathbf{j}(\mathbf{x},t)
    \end{split}
    \label{seq:fp2}
\end{equation}
where, $\mathbf{j}(\mathbf{x},t) \equiv (-\mathbf{F}\mathbf{x}P({\mathbf{x},t}) - \mathbf{D} \nabla P(\mathbf{x},t))$  denotes the probability current in the phase space.

Starting from an arbitrary initial condition for $\mathbf{x}$, the system will reach a non-equilibrium steady state in the long time ($\lim t \rightarrow \infty$) with a characteristic probability distribution and current given by,
\begin{align}
\begin{split}
    P_{ss}(\mathbf{x}) = ((2\pi)^{3/2} \sqrt{\det \mathbf{ C}})^{-1} e^{-\frac{1}{2}\mathbf{x}^T\mathbf{C}^{-1}\mathbf{x}} \\
    \mathbf{j}_{ss}(\mathbf{x}) =  (-\mathbf{F}\mathbf{x} + \mathbf{D}\mathbf{C}^{-1}\mathbf{x})P_{ss}(\mathbf{x}),
    \end{split}
    \label{seq:rho_current2}
\end{align}
in terms of the long-time limit of the covariance matrix $\mathbf{C}(t)$. \\
\textit{Covariance matrix.-} Note that the diffusion tensor is not diagonal in form, making the  system non-multipartite \cite{leighton2024jensen}. Following the technique introduced in \cite{kwon2011nonequilibrium}, the steady-state covariance matrix $\mathbf{C}$ is given by 
\begin{equation}
    \mathbf{C} = \mathbf{F}^{-1}(\mathbf{D} + \mathbf{Q}),
\end{equation}
where $\mathbf{Q}$ is an antisymmetric $3 \times 3$ matrix that can be uniquely determined by
\begin{equation}
    \mathbf{FQ} + \mathbf{QF}^T = \mathbf{FD} - \mathbf{DF}^T.
\end{equation}
%If $\mathbf{Q}$ is nonzero in the NESS, it implies the violation of the detailed balance in the system.
The elements of the steady state ($\lim t \rightarrow \infty$) covariance matrix for this system are,
\begin{align}
\label{eq:cov_ou}
\begin{split}
    & C_{11} \equiv \langle x_1 x_1 \rangle = \frac{D_0 \gamma}{k_1} +\frac{D_e k_1(\gamma(k_1 +k_2 -\epsilon^2 k_2)+(1-\epsilon^2)k_2(k_1 +k_2)\tau_e)}{(k_1 +k_2)(\gamma^2 + \gamma(k_1 +k_2)\tau_e + (1-\epsilon^2)k_1 k_2 \tau_e^2)},\\
    & C_{12} \equiv \langle x_1 x_2 \rangle = \frac{\epsilon D_e  \gamma k_1^2}{(k_1+k_2)(\gamma^2 + \gamma(k_1 + k_2)\tau_e + (1-\epsilon^2)k_1 k_2 \tau_e^2)},\\
    & C_{13} \equiv \langle x_1 \lambda\rangle = \frac{D_e k_1(\gamma + (1-\epsilon^2)k_2 \tau_e)}{(\gamma^2 + \gamma(k_1 + k_2)\tau_e + (1-\epsilon^2)k_1 k_2 \tau_e^2)},\\
    & C_{21} \equiv \langle x_2 x_1 \rangle = C_{12},\\
    & C_{22} \equiv \langle x_2 x_2 \rangle = \frac{D_0 \gamma}{k_2} + \frac{\epsilon^2 D_e \gamma k_1^2}{(k_1+k_2)(\gamma^2 + \gamma(k_1 + k_2)\tau_e + (1-\epsilon^2)k_1 k_2 \tau_e^2)},\\
    & C_{23} \equiv \langle x_2 \lambda\rangle = \frac{\epsilon D_e \gamma k_1}{\gamma^2 + \gamma(k_1+k_2)\tau_e + (1-\epsilon^2)k_1 k_2 \tau_e^2},\\
    & C_{31} \equiv \langle \lambda x_1\rangle = C_{13},\\
    & C_{32} \equiv \langle \lambda x_2 \rangle = C_{23},\\
    & C_{33} \equiv \langle \lambda \lambda \rangle = \frac{D_e}{\tau_e}.
    \end{split}
\end{align}
If the particles are well separated such that hydrodynamic coupling is negligible ($\epsilon \rightarrow 0$), the covariance matrix is,
\begin{equation}
\label{eq:cov_ep0}
    \mathbf{C}|_{\epsilon \rightarrow 0}=
    \begin{pmatrix}
    \frac{D_0 \gamma}{k_1} + \frac{D_e k_1}{\gamma + k_1 \tau_e} & 0 & \frac{D_e k_1}{\gamma + k_1 \tau_e} \\
    0 & \frac{D_0 \gamma}{k_2} & 0\\
    \frac{D_e k_1}{\gamma + k_1 \tau_e} & 0 & \frac{D_e}{\tau_e}
    \end{pmatrix}
\end{equation}

\textit{Stationary state of the two-temperature model}: We can find the covariance matrix corresponding to modulation by a Gaussian white noise by taking $\tau_e \rightarrow 0$ (with $D_e \neq 0$) in Eq.\eqref{eq:cov_ou}. This gives,
\begin{equation}
\label{eq:cov_w}
    \begin{split}
        & [C_{11}]_{tw} = \frac{D_0 \gamma}{k_1} + \frac{D_e k_1^2}{\gamma k_1} -\epsilon^2\frac{k_2}{k_1}\frac{D_e k_1^2}{\gamma(k_1 + k_2)},\\
        & [C_{12}]_{tw} = \epsilon \frac{D_e k_1^2}{\gamma(k_1 + k_2)},\\
        & [C_{22}]_{tw} =  \frac{D_0 \gamma}{k_2} + \epsilon^2 \frac{D_e k_1^2}{\gamma(k_1 + k_2)}.
    \end{split}
\end{equation}
If we consider $D_e k_1^2 = k_B\Delta T \gamma$ to match the noise correlation of the external Gaussian modulation of Ref. \cite{berut2014energy}, Eqs.\eqref{eq:cov_w} will be transformed to,
\begin{equation}
\label{eq:cov_w_berut}
    \begin{split}
        & [C_{11}]_{tw} = \frac{k_B T}{k_1} + \frac{k_B \Delta T}{k_1} -\epsilon^2\frac{k_2}{k_1}\frac{k_B \Delta T}{(k_1 + k_2)},\\
        & [C_{12}]_{tw} = \epsilon \frac{k_B \Delta T}{(k_1 + k_2)},\\
        & [C_{22}]_{tw} =  \frac{k_B T}{k_2} + \epsilon^2 \frac{k_B \Delta T}{(k_1 + k_2)},
    \end{split}
\end{equation}
which match exactly with the Eqs.(14) of Ref.\cite{berut2014energy}. Moreover, Eq.~\eqref{eq:cov_w_berut} are the elements of the steady state covariance matrix of the \textit{`two-temperature model'} we discussed in the main text. In the experimental studies of this model~\cite{berut2014energy,berut2016stationary}, the two-temperature configuration was created by forcing one of the particles by random white noise while the other particle was kept in close proximity. Note that, the effect of the random white noise is assimilated as an `effective temperature' which is attributed to the different temperatures of the two particles. The details of the external driving (trajectory) are not explicitly available as a degree of freedom, unlike in the \textit{OU-noise driven model}. \\

It can be interesting to point out that the steady-state cross-variances of the hydrodynamically coupled system (with two microparticles) will be non-zero only in the out-of-equilibrium configuration. In an equilibrium situation ($D_e = 0$ or $\Delta T =0$), the two particles will be statistically independent with the variance $\frac{k_B T}{k_1}$ and $\frac{k_B T}{k_2}$ respectively.

\subsection{ Entropy production rate and Short-time inference technique}
\label{ap2:inference_tech}
In this section, we briefly describe the inference technique to estimate the entropy production rate from trajectory data.\\

 The trajectory-dependent entropy for a nonequilibrium system is defined as~\cite{seifert2005entropy}, 
 \begin{equation}
     \label{eq:trajentdef}
     \pi(t) = - \ln P(\mathbf{x},t),
 \end{equation}
 where $P(\mathbf{x},t)$ is solution of the Fokker-plank equation, Eq.~\eqref{seq:fp2}. The rate of change of entropy (Eq.~\eqref{eq:trajentdef}) is then
\begin{equation}
\label{eq:trajentratedef}
    \dot{\pi}(t) = -\frac{\partial_t P(\mathbf{x},t)}{P(\mathbf{x},t)}\Bigg\vert_{\mathbf{x}(t)} - \frac{\nabla_\mathbf{x} P(\mathbf{x},t)}{P(\mathbf{x},t)}\Bigg\vert_{\mathbf{x}(t)} \dot{\mathbf{x}},
\end{equation}
which can be further reduced to,
\begin{equation}
\label{eq:trajentratedef2}
    \dot{\pi}(t) = -\frac{\partial_t P(\mathbf{x},t)}{P(\mathbf{x},t)}\Bigg\vert_{\mathbf{x}(t)} +\frac{\mathbf{D}^{-1}\mathbf{j}({\mathbf{x},t}) }{ P(\mathbf{x},t)}\Bigg\vert_{\mathbf{x}(t)} \dot{\mathbf{x}} - \mathbf{D}^{-1}\mathbf{Fx}\Big\vert_{\mathbf{x}(t)} \dot{\mathbf{x}},
\end{equation}
using the definition of nonequilibrium current of the Fokker-plank equation(Eq.~\eqref{seq:fp2}) as shown in Ref.~\cite{seifert2005entropy}. Now the third term of the Eq.~\eqref{eq:trajentratedef2} can be related to the rate of heat flow into the medium and can be identified as an increase in entropy of the medium $\pi_m$ as $\dot{Q}(t) = \mathbf{D}^{-1}\mathbf{Fx}\dot{\mathbf{x}} \equiv T \dot{\pi}_m(t)$. \\
Then Eq.~\eqref{eq:trajentratedef2} can be written as,
\begin{align}
\begin{split}
    \dot{\pi}_{tot}(t) &= \dot{\pi}_m(t) + \dot{\pi}(t)\\
    &= -\frac{\partial_t P(\mathbf{x},t)}{P(\mathbf{x},t)}\Bigg\vert_{\mathbf{x}(t)} +\frac{\mathbf{D}^{-1} \mathbf{j}({\mathbf{x},t}) }{P(\mathbf{x},t)}\Bigg\vert_{\mathbf{x}(t)}\dot{\mathbf{x}}
    \end{split}
\end{align} 
Finally, the average rate of  total entropy production can be estimated as~\cite{seifert2005entropy},
\begin{equation}
    \Pi = \langle \dot{\pi}_{tot}(t)\rangle = \int d\mathbf{x} \frac{\mathbf{D}^{-1} \mathbf{j}^2(\mathbf{x},t)}{P(\mathbf{x},t)},
\end{equation}
since averaging over all the trajectories at a given $(\mathbf{x},t)$ leads to, $\langle \dot{\mathbf{x}}|\mathbf{x},t\rangle = \mathbf{j}(\mathbf{x},t)/P(\mathbf{x},t) $ and upon averaging over $\mathbf{x}$, $\int \ d\mathbf{x} \partial_t P(\mathbf{x},t) = 0$.

 Now, the average entropy production rate for a nonequilibrium system in a steady state with the  probability density $\rho_{ss}(\mathbf{x})$ and current $\mathbf{j}_{ss}(\mathbf{x})$  can be written as 
\begin{equation}
    \Pi = \int d\mathbf{x} \frac{\mathbf{D}^{-1} \mathbf{j}_{ss}^2(\mathbf{x})}{P_{ss}(\mathbf{x})}.
    \label{ap:eq_epr}
\end{equation}
Here we use this formula to compute the analytical expressions for the entropy production rate for both models.
Even though, the entropy production rate of a non-equilibrium system can be analytically estimated using this methodology, directly applying these equations to experimental data offers a number of practical challenges in a generic setting. The difficulties range from identifying the right conversion factors in experiments to accurately being able to reproduce the Langevin model in an experiment. We elaborated on some of these challenges in an earlier work~\cite{manikandan2021quantitative}. Here, motivated by our prior works in this direction, we have used an indirect estimation technique - referred to as the short-time inference scheme -  for computing the entropy production rate, which is based on the thermodynamic uncertainty relation~\cite{barato2015thermodynamic}. This method enables the computation of entropy production rate in a data-driven manner, and is agnostic of the specific details of the model (for example the parameter values) or the conversion factors. The short-time inference technique was first introduced for overdamped  $Langevin$ dynamics in Ref. \cite{manikandan2020inferring} and rigorously proved in Refs. \cite{van2020entropy,otsubo2020estimating}. It was subsequently used to estimate the entropy production rate from experimental trajectories in Ref. \cite{manikandan2021quantitative} where it was also compared with analytically known expressions. Moreover, this technique is also applicable to systems with non-linear potential landscapes \cite{das2022inferring} and to inferring the activity of biological cell membranes \cite{manikandan2024estimate}.

Using this technique, we can estimate the entropy production rate of a system in a non-equilibrium steady state as
\begin{equation}
    \Pi = \lim_{dt \rightarrow 0} \max_{J} \left[\frac{2 k_B \langle J \rangle^2}{dt \ \text{Var}(J)}\right],
    \label{eq:ap_TUR_J}
\end{equation}
where $J$ is a weighted scalar current - that can be computed from the time discretized experimental or numerical trajectory data ($\mathbf{x}^{i}$) sampled at an interval of $dt$ - as
\begin{equation}
    J = \mathbf{f}\left(\frac{\mathbf{x}^{i+1} + \mathbf{x}^{i} }{2}\right)\cdot\left(\mathbf{x}^{i+1} -\mathbf{x}^{i}\right).
    \label{eq:ap_J}
\end{equation}
Motivated by the linearity of the systems, we have approximated 
$\mathbf{f(x)}$ as the linear combination of linear basis functions $\psi_m(\mathbf{x})$ of different dimensions (equal to the number of degrees of freedom considered in a model) as $\mathbf{f(x)} = \sum_{m=1}^M \mathbf{a}_{m}\psi_m(\mathbf{x})$. For any such adequate representation of $\mathbf{f(x)}$, an analytical solution to the maximisation problem of Eq.\eqref{eq:ap_TUR_J} is known \cite{van2020entropy} and given by,
\begin{equation}
    \label{ap:eq_opt}
    \Pi = \frac{2 \langle \psi_k \rangle (\boldsymbol{\Xi}^{-1})_{k,l} \langle \psi_l \rangle}{dt},
\end{equation}
where, $(\boldsymbol{\Xi})_{k,l} = \langle \psi_k \psi_l \rangle - \langle \psi_k \rangle \langle \psi_l \rangle $. 
To find more details about the method, we refer to  Sec. II B of Ref. \cite{van2020entropy} and to  Sec. III of Ref. \cite{das2022inferring}. \\

\subsection{ Partial entropy production rate of \textit{OU-noise driven} model}
\label{ap3:partial_ou}
In this section, we provide a detailed analytical estimation of the partial entropy production rate of the \textit{OU-noise driven} model when $\lambda(t)$ is intractable or hidden.\\

\begin{figure}[h]
    \centering
    \includegraphics[width=0.99\linewidth]{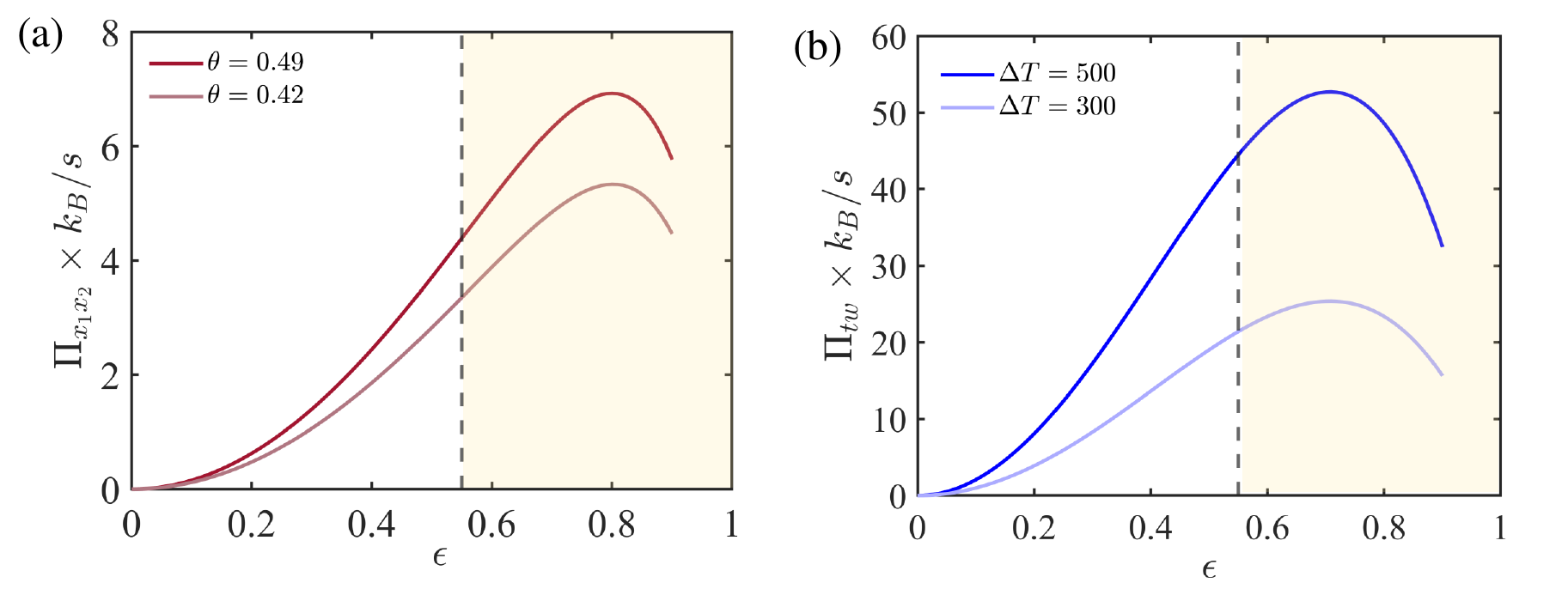}
    \caption{Functional dependence of (a) $\Pi_{x_1 x_2}$ (Eq.\eqref{ap:eq_reduced_epr}) and (b) $\Pi_{tw}$ (Eq.~\eqref{eq:epr_w} of main text) on the parameter $\epsilon$. Both quantities show similar non-monotonic behaviour with $\epsilon$. However, achieving hydrodynamic interaction strength beyond $\epsilon \sim 0.55$, will not be physically possible (shaded region of the plots) in this system as the interparticle separation needs to be extremely low. 
    Moreover, for such low interparticle separation ($d$) such that $(d-2a)/2a<<1$, the functional form of hydrodynamic interaction strength ($\epsilon$) with $d$ will be very different such that the Langevin equation may not be valid either.}
    \label{sfig:epr_reduced_two}
\end{figure}
The expression to find the total entropy production rate ( Eq.~\eqref{ap:eq_epr}) can be rewritten in terms of the drift ($\mathbf{F}$), diffusion ($\mathbf{D}$) and covariance matrix ($\mathbf{C}$) as shown in Ref.~\cite{nicoletti2024tuning}, 
\begin{equation}
\label{ap:eq_epr_matrix}
    \Pi = \text{Tr}\ [\mathbf{D}^{-1}\mathbf{F}\mathbf{C}\mathbf{F}^{T}] - \text{Tr}\ [\mathbf{F}], 
\end{equation} 
which is useful to estimate the partial entropy production rate of the system based on the available degrees of freedom. According to the technique introduced in Ref.~\cite{nicoletti2024tuning}, the partial entropy production rate of a system with only $x_1(t)$ and $x_2(t)$ as available degrees of freedom can be computed using Eq.~\eqref{ap:eq_epr_matrix} by substituting the diffusion matrix $\mathbf{D}$ and covariance matrix ($\mathbf{C}$) by their corresponding submatrices $\mathbf{D}_{x_1 x_2}$ and $\mathbf{C}_{x_1 x_2}$.  The drift matrix $\mathbf{F}$ however needs to be redefined as a reduced drift matrix $\mathbf{F}^{red}_{x_1 x_2}$ such that, 
\begin{equation}
    \mathbf{F}^{red}_{x_1 x_2} = \mathbf{F}_{x_1 x_2} + \mathbf{B}\mathbf{C}_{x_1 x_2}^{-1}, \\\\\\\\\ 
       (\mathbf{B})_{ij} = (\mathbf{F})_{i\lambda} (\mathbf{C})_{j\lambda}
    \label{ap:eq_reduced_F}
\end{equation}
with $i, j = x_1, x_2$. For the \textit{`OU-noise driven system'}, the relevant submatrices corresponding to the degrees of freedom $x_1$ and $x_2$ are,  
\begin{equation}
\label{ap:eq_submatrix}
\mathbf{F}_{x_1 x_2} = \begin{pmatrix}
    k_1/\gamma & \epsilon k_2/\gamma\\
    \epsilon k_1/\gamma & k_2/\gamma
\end{pmatrix},\\\
\mathbf{D}_{x_1 x_2} = \begin{pmatrix}
    D_0 & \epsilon D_0\\
    \epsilon D_0 & D_0
\end{pmatrix}, \\\
\mathbf{C}_{x_1 x_2} = \begin{pmatrix}
    C_{11} &  C_{12}\\
     C_{21} & C_{22}
\end{pmatrix},
    \end{equation}
    where $C_{11}$, $C_{12}$, $C_{21}$ and $C_{22}$ are given in Eq.~\eqref{eq:cov_ou}. The matrix $\mathbf{B}$ can be computed as, 
    \begin{equation}
        \mathbf{B} = \begin{pmatrix}
            -\frac{k_1^2 D_e \left(\gamma -k_2 \left(\epsilon ^2-1\right) \tau _e\right)}{\gamma  \left(\gamma ^2+\tau _e \left(k_1 \left(\gamma -k_2 \left(\epsilon ^2-1\right) \tau _e\right)+\gamma  k_2\right)\right)} & -\frac{k_1^2 \epsilon  D_e}{\gamma ^2+\tau _e \left(k_1 \left(\gamma -k_2 \left(\epsilon ^2-1\right) \tau _e\right)+\gamma  k_2\right)}\\
            -\frac{k_1^2 \epsilon  D_e \left(\gamma -k_2 \left(\epsilon ^2-1\right) \tau _e\right)}{\gamma  \left(\gamma ^2+\tau _e \left(k_1 \left(\gamma -k_2 \left(\epsilon ^2-1\right) \tau _e\right)+\gamma  k_2\right)\right)} & -\frac{k_1^2 \epsilon ^2 D_e}{\gamma ^2+\tau _e \left(k_1 \left(\gamma -k_2 \left(\epsilon ^2-1\right) \tau _e\right)+\gamma  k_2\right)}
        \end{pmatrix}
    \end{equation}.

Using these matrices, the reduced drift matrix of the system can be computed and the partial entropy production rate of the system can be subsequently estimated as,
\begin{equation}
    \label{ap:eq_reduced_epr}
    \Pi_{x_1 x_2} = \frac{k_1^4 k_2^2 \epsilon ^2 \left(1-\epsilon ^2\right) D_e^2 \left(\gamma  D_0 \left(k_1+k_2\right) \chi -k_1^2 k_2 \left(\epsilon ^2-1\right) D_e \left(\gamma +k_2 \tau _e\right)\right)}{\gamma  D_0 \chi  \left(D_0 \left(k_1+k_2\right){}^2 k_1^2 \chi  D_e \left(\gamma -k_2 \left(\epsilon ^2-1\right) \tau _e\right)+k_2^2 k_1^4 \left(-\epsilon ^2\right) \left(\epsilon ^2-1\right) D_e^2 \left(\gamma +\left(k_1+k_2\right) \tau _e\right)+\gamma  D_0^2 \left(k_1+k_2\right){}^2 \chi ^2\right)},
\end{equation}
with $\chi = \gamma ^2+\tau _e \left(k_1 \left(\gamma -k_2 \left(\epsilon ^2-1\right) \tau _e\right)+\gamma  k_2\right)$.

The partial entropy production rate ($\Pi_{x_1x_2}$) is found to increase as hydrodynamic interaction strength ($\epsilon$) is enhanced. It also vanishes as $\epsilon\rightarrow0$, similar to the entropy production rate of the \textit{two-temperature model} ($\Pi_{tw}$) as shown in Fig.~\ref{sfig:epr_reduced_two}.

\subsection{ Validation of thermodynamic laws}
\label{ap4:energetics}

In the presence of hydrodynamic interactions, the noises for the different degrees of freedom are cross-correlated
and it is not completely obvious how to get quantities such as heat or work unambiguously from the trajectories~\cite{leighton2024jensen}.
Previous studies have suggested that energy balance does not hold at the level of individual trajectories or at the level of ensemble averages, for systems having hydrodynamic interactions ~\cite{berut2014energy,berut2016stationary,berut2016theoretical}. In the following, we argue that this claim is incorrect and that the degrees of freedom of the system account for all the energetics involved. First, using the `OU-noise driven model' we show that the presence of hydrodynamic interactions alone does not cause violations of the energy balance conditions if the thermodynamic quantities, especially the dissipated heat, are identified correctly. We also show that previous prescriptions from Refs. ~\cite{berut2014energy,berut2016stationary,berut2016theoretical} lead to inconsistencies in identifying the medium entropy production.
\subsection*{OU-noise driven model} 

For the \textit{`OU-noise driven model'},
the total potential energy of the system is,
\begin{align}
V(x_1,x_2,\lambda ) = \frac{k_1}{2} (x_1-\lambda)^2     + \frac{k_2}{2} x_2^2  
\end{align}
The change in energy of the system during a process of duration $\tau$ is,
\begin{align}
    \Delta E = V(x_1(\tau),x_2(\tau),\lambda (\tau)) -  V(x_1(0),x_2(0),\lambda (0)).
\end{align}
The work done is,
\begin{align}
 \begin{split}
        W_\tau &= \int_0^\tau dt\; \partial_\lambda  V(x_1,x_2,\lambda ) \circ\dot{\lambda}\\
        &= k_1\int_0^\tau dt\;  (\lambda - x_1)\circ\dot{\lambda}.
 \end{split}
\end{align}
Similarly, the definition of the dissipated heat following the standard definition \cite{seifert2012stochastic} should be:
\begin{align}
 \begin{split}
        Q_\tau
        &=  \int_0^\tau dt\; \partial_{\bm x}  V(x_1,x_2,\lambda ) \circ \dot{\bm x}\\
        &=  \int_0^\tau dt\; \partial_{ x_1}  V(x_1,x_2,\lambda ) \circ \dot{ x}_1 +  \int_0^\tau dt\; \partial_{ x_2}  V(x_1,x_2,\lambda ) \circ \dot{ x}_2 \equiv Q_{1\tau} +Q_{2\tau}
 \end{split}
 \label{eq:heatdefn}
 \end{align}
 with $\bm x = (x_1,x_2)$. \\ 
From the above two equations, it follows that, 
\begin{align}
\label{ap:eq_heat_c}
\begin{split}
       Q_\tau + W_\tau &=  \int_0^\tau dt\; \partial_{\bm x}  V(x_1,x_2,\lambda ) \circ \dot{\bm x} + \int_0^\tau dt\; \partial_\lambda  V(x_1,x_2,\lambda )\circ \dot{\lambda}\\
       &=V(x_1(\tau),x_2(\tau),\lambda (\tau)) -  V(x_1(0),x_2(0),\lambda (0))\\
       &=\Delta E.
\end{split}
\end{align}
This establishes energy balance at the level of individual trajectories for the `OU-noise driven model' just using the standard notions of heat and work. Crucially, we note that the presence of hydrodynamic interactions does not affect the energy balance condition. 

Since energy balance holds at the level of individual trajectories, it must hold at the level of ensemble averages as well. To demonstrate this, we first note that at the ensemble level, the input power i.e. the rate of average work (in units of $k_B T/s$) is \footnote{
We can explicitly calculate the averages following the method shown in Ref. \cite{seifert2005entropy,seifert2012stochastic} as
$
\langle f(x) \circ  \dot{x} \rangle = \int dx f(x)j_{ss}(x). 
$
}, from Eq. (29),
\begin{equation}
\begin{split}
\label{eq:input_power}
        \langle\dot{W}\rangle &= k_1 \langle (\lambda(t) - x_1(t)) \circ \dot{\lambda}(t) \rangle\\
        &= k_1 \langle \lambda(t) \circ \dot{x}_1(t)  \rangle \\
        &= \frac{D_e k_1^2 (\gamma +(1-\epsilon^2) k_2 \tau_e)}{ \tau_e(\gamma^2 + \gamma(k_1 +k_2)\tau_e + (1-\epsilon^2)k_1k_2\tau_e^2)}.
\end{split}
\end{equation}
Since $\langle \Delta E \rangle = 0$ in the stationary state, if energy is conserved, we expect Eq. (\ref{eq:input_power}) to exactly equal the average total heat dissipated per unit time into the environment. In addition, since the system is isothermal, 
 $ \langle \Delta \dot{S}_{m} \rangle = - \frac{\langle \dot{Q} \rangle}{T}$. 

First, we obtain $\langle\dot{Q}\rangle $
 from an ensemble average of  Eq.~\eqref{eq:heatdefn} as,
        \begin{align}
        \label{eq:Qav}
            \begin{split}
                \langle\dot{Q}\rangle & = \langle \partial_{\bm x}  V(x_1,x_2,\lambda ) \circ \dot{\bm x} \rangle\\
                & = \langle k_1(x_1 - \lambda) \circ \dot{x}_1 \rangle + \langle k_2 x_2\circ\dot{x}_2\rangle\\
                &= - \frac{D_e k_1^2 (\gamma +(1-\epsilon^2) k_2 \tau_e)}{ \tau_e(\gamma^2 + \gamma(k_1 +k_2)\tau_e + (1-\epsilon^2)k_1k_2\tau_e^2)},
            \end{split}
        \end{align}
Clearly the average dissipated heat is related to the average rate of work done given in Eq.~\eqref{eq:input_power} as $-\langle\dot{Q}\rangle = \langle\dot{W}\rangle$. Hence we can establish energy balance at the level of ensemble averages too as  $\langle\dot{Q}\rangle + \langle\dot{W}\rangle = 0$.\\ 

\noindent
\textbf{Identifying heat in terms of medium entropy production}: A more general approach to identifying the functional form of the total dissipated heat is to first identify the medium entropy production rate. 
For a generic Langevin equation of the form,
      \begin{align}
          \dot{\bm x} = -\mathcal{\bm F}({\bm x}(t)) + {\bm \xi}(t),
      \end{align}
      where $\langle {\bm \xi}(t) {\bm \xi}(s) = 2{\bm D} \delta (t-s)$, the medium entropy production can always be identified along an individual trajectory as \cite{seifert2012stochastic},
      \begin{align}
      \label{eq:med_ep}
          \Delta \dot{S}_{m} = k_B \int_0^\tau {\bm D}^{-1} \mathcal{\bm F}({\bm x}(t)) \circ d{\bm x}(t).
      \end{align}
      Eq.\ \eqref{eq:med_ep} follows from the definition \cite{seifert2005entropy,seifert2012stochastic},
      \begin{align}
          \Delta \dot{S}_{m} = k_B \ln \frac{P[{\bm x}(\cdot) \vert {\bm x}_0]}{P[\tilde{\bm x}(\cdot) \vert \tilde{\bm x}_0]},
      \end{align}
    where $P[{\bm x}(\cdot) \vert {\bm x}_0]$ is the conditional path probability of the trajectory ${\bm x}(\cdot)$ defined as,
    \begin{align}
        P[{\bm x}(\cdot) \vert {\bm x}_0] \propto \exp \left(-\int_0^\tau dt \left(\dot{\bm x}  - \mathcal{\bm F}({\bm x}(t)) \right)^T (2{\bm D} )^{-1} \left(\dot{\bm x}  - \mathcal{\bm F}({\bm x}(t)) \right) \right) 
    \end{align}
    and $\tilde{\bm x}(t) \equiv  {\bm x}(\tau -t)$ is the corresponding time - reversed trajectory. The total entropy production is then given by, $\Delta \dot{S}_{tot} =  \Delta \dot{S}_{m} + \Delta \dot{S}_{sys}$, where $\Delta \dot{S}_{sys} = {\rm ln} \frac{p[ {\bm x}_0]}{p[ \tilde{\bm x}_0]}$.  Note that for an isothermal system, $\Delta \dot{S}_{m} = - Q/T$, where $Q$ is the total heat dissipated to the reservoir at temperature $T$.  For the \textit{OU-noise driven} model, this approach leads to the identification,
\begin{equation}
\begin{split}
\label{eq:mediumep_path}
   \frac{\langle\dot{Q}\rangle}{T}  =& \langle \Delta \dot{S}_{m}\rangle \equiv k_B\ \langle \mathbf{D}^{-1} \mathbf{(Fx) \circ \dot{x}}\rangle\\
   \implies \langle \Delta \dot{S}_{m}\rangle =& \ k_B\int d\mathbf{x} \ \mathbf{D}^{-1} \mathbf{(F x)}\mathbf{j}_{ss}(\mathbf{x}), 
\end{split}
\end{equation}
with $\mathbf{j}_{ss}(\mathbf{x}) =  (-\mathbf{F}\mathbf{x} + \mathbf{D}\mathbf{C}^{-1}\mathbf{x})P_{ss}(\mathbf{x})$. 
The integral is then estimated to be, 
\begin{equation}
\label{eq:mediumep_path_f}
    \begin{split}
     \langle \Delta \dot{S}_{m}\rangle  =& -k_B\frac{D_e k_1^2 (\gamma +(1-\epsilon^2) k_2 \tau_e)}{D_0 \gamma \tau_e(\gamma^2 + \gamma(k_1 +k_2)\tau_e + (1-\epsilon^2)k_1k_2\tau_e^2)},\\
   \implies \langle\dot{Q}\rangle =& D_0 \gamma 
    \langle \Delta \dot{S}_{m}\rangle/k_B
    \\
   =& -\frac{D_e k_1^2 (\gamma +(1-\epsilon^2) k_2 \tau_e)}{ \tau_e(\gamma^2 + \gamma(k_1 +k_2)\tau_e + (1-\epsilon^2)k_1k_2\tau_e^2)}.
\end{split}
\end{equation}
Note that, as expected, this expression is identical to the expression obtained 
 from an ensemble average of  Eq.~\eqref{eq:heatdefn} obtained in Eq.\ \eqref{eq:Qav}. The formula above also correctly identifies heat along individual trajectories in Eq.\ \eqref{eq:heatdefn}. Note that Eq.\ \eqref{eq:mediumep_path} will give a term corresponding to the degree of freedom $\lambda$ as well. But this term is annulled by it's own system entropy contribution as it should, since $\lambda(t)$
 by itself follows an equilibrium process and does not contribute to the total entropy production.
        
However, if we use the definitions in Refs.~\cite{berut2014energy,berut2016stationary,berut2016theoretical} for dissipated heat, then we run into several inconsistencies. There, the heat fluxes due to the individual particles are defined as $Q_i(\tau) = \int_0^{\tau} dt \ (\gamma \dot{x}_i - \eta_i)\circ \dot{x}_i $. This definition neglects the fact that the Diffusion matrix is non-diagonal. Nevertheless, if we accept such a definition, the amount of heat released by the 1st particle (externally perturbed) during the time $\tau$ is,\ $
        Q_{1\tau} = \int_0^\tau dt\ (\gamma \dot{x}_1(t) - \eta_1(t))\circ \dot{x}_1(t)
        = \int_0^\tau dt\ (- k_1 [x_1(t) - \lambda(t)]  -\epsilon k_2 x_2(t)) \circ \dot{x}_1(t)$. Similarly, the heat dissipated by the 2nd particle is defined as $Q_{2\tau} = \int_0^\tau dt\ (\gamma \dot{x}_2(t) - \eta_2(t))\circ \dot{x}_2(t) 
        = \int_0^\tau dt\ (-\epsilon k_1 [x_1(t) - \lambda(t)]  - k_2 x_2(t)) \circ \dot{x}_2(t)$. Using these definitions, the total rate of heat dissipated is estimated as,
        \begin{equation}
        \label{ap:eq_heat_rate_diff}
            \begin{split}
                \langle\dot{Q}\rangle &= \langle\dot{Q}_1\rangle + \langle\dot{Q}_2\rangle \\
                & = \frac{k_1^2 D_e \left(k_2 \left(\gamma  \left(\epsilon ^2+1\right)+k_2 \left(\epsilon ^2-1\right)^2 \tau _e\right)+k_1 \left(\gamma  \left(\epsilon ^2+1\right)-k_2 \left(\epsilon ^4-1\right) \tau _e\right)\right)}{\left(k_1+k_2\right) \tau _e \left(\gamma ^2+\tau _e \left(k_1 \left(\gamma -k_2 \left(\epsilon ^2-1\right) \tau _e\right)+\gamma  k_2\right)\right)},
            \end{split}
        \end{equation}
        which is evidently different from the average rate of work computed in Eq.~\eqref{eq:input_power}.  This definition of total dissipated heat also does not satisfy the condition $ \langle \Delta \dot{S}_{m} \rangle = - \frac{\langle \dot{Q} \rangle}{T}$, given the standard definition of medium entropy as in Eq. (\ref{eq:mediumep_path}).

\subsection*{Two-temperature model}

For the `\textit{two-temperature model'} studied in Refs. \cite{berut2014energy,berut2016stationary,berut2016theoretical}, as mentioned in the previous section, the heat fluxes due to the respective particles was estimated as $Q_i(\tau) = \int_0^{\tau} dt \ (\gamma \dot{x}_i - \eta_i)\circ \dot{x}_i $.  These studies showed that the heat released from the `hot' particle was not fully exhausted by the `cold' one if the particles were trapped in potentials with different stiffnesses  - leading to the `apparent' violation of the energy balance relation.  It was also argued that energy would not be conserved for this system due to the dissipative nature of the coupling and energy conservation could be restored if the two particles are trapped in potentials with equal stiffnesses as the system will then be indistinguishable from those with conservative coupling \cite{berut2016theoretical}. 

As we explained in the previous Section, the presence of hydrodynamic interactions does not cause the violation of the energy balance condition. So in the following, we first assume that energy balance should hold, and use that to identify $Q_{1,2}(\tau)$. Note that, for the \textit{two-temperature model}, there is no explicit external driving, so $W = 0$.  Assuming that the energy is conserved, we can then write,\  $Q_{1\tau} + Q_{2\tau} = V(x_1(\tau),x_2(\tau)) - V(x_1(0),x_2(0))$ where $V(x_1,x_2) = 1/2 (k_1 x_1^2 + k_2 x_2^2)$. From this, we get, $Q_{2\tau}  = V(x_1(\tau),x_2(\tau)) - V(x_1(0),x_2(0)) -Q_{1\tau}$.
        Now in order to identify $Q_{1\tau}$ and $Q_{2\tau}$ at the level of individual trajectories, we first identify the medium entropy production rate in this system. 
        This can be easily computed from the standard definition~\cite{seifert2005entropy},
         \begin{align}
            \begin{split}
                \Delta S_{m(\tau)}^{tw} &\equiv k_B \int_0^\tau dt\ \mathbf{D}_{tw}^{-1} (\mathbf{F}_{tw}\mathbf{x}) \circ \dot{\mathbf{x}}\\ &=  \int_0^\tau dt\ \frac{ (k_1 T x_1-\Delta T k_2 x_2 \epsilon )\circ \dot{x}_1+k_2 (\Delta T+T)x_2 \circ \dot{x}_2}{T (\Delta T+T)} 
            \end{split}
        \end{align}
        Now using the standard definition of medium entropy production for a two - temperature system, we can write, 
        \begin{align}
\begin{split}
                \Delta S_{m(\tau)}^{tw} &= -\frac{Q_{1\tau}}{T} - \frac{Q_{2\tau}}{T + \Delta T}\\ 
                &= -\frac{Q_{1\tau}}{T} - \frac{V(x_1(\tau),x_2(\tau)) - V(x_1(0),x_2(0)) -Q_{1\tau}}{T + \Delta T}
\end{split}
\label{eq:SQrel_two}
        \end{align}
        In the second equation of Eq.~\eqref{eq:SQrel_two}, we have applied the energy balance condition in rewriting $Q_{2\tau}$. Equating the two, we can identify
        \begin{align}
        \label{eq:SQ1}
     Q_{1\tau} =\int_0^\tau dt\ \frac{-(V(x_1(\tau),x_2(\tau)) - V(x_1(0),x_2(0))) T+(\Delta T k_2 x_2 \epsilon -k_1 T x_1)\circ \dot{x}_1 -k_2 (\Delta T+T)x_2\circ \dot{x}_2}{\Delta T}
\end{align}
        and
       \begin{align}
       \label{eq:SQ2}
\begin{split}
                Q_{2\tau} &= V(x_1(\tau),x_2(\tau)) - V(x_1(0),x_2(0)) -Q_{1\tau} \\
                &=\int_0^\tau dt\ \frac{(\Delta T+T) \left((V(x_1(\tau),x_2(\tau)) - V(x_1(0),x_2(0)))+k_2x_2 \circ\dot{x}_2\right)+ (k_1T x_1-\Delta T k_2x_2 \epsilon )\circ\dot{x}_1}{\Delta T}.
\end{split}
        \end{align}
We can then estimate the average rate of heat flow at steady state as,
\begin{align}
    \begin{split}
        -\langle \dot{Q}_1 \rangle &= \langle \dot{Q}_2 \rangle = - \langle \epsilon k_2 x_2 \circ \dot{x}_1 \rangle \\
        &= - \frac{k_2^2  \epsilon^2 (1-\epsilon^2)}{(k_1 + k_2) \gamma}k_B \Delta T,
    \end{split}
\end{align}
since in a steady state, $\langle V(x_1(\tau),x_2(\tau)) - V(x_1(0),x_2(0))\rangle = 0$ and $\langle x_1 \circ \dot{x}_1 \rangle = \langle x_2 \circ \dot{x}_2 \rangle = 0$.

The same expression as above was identified as the heat flow from hot particle to cold particle in Refs.~\cite{berut2016stationary,berut2016theoretical}. We argue that it needs to be entirely associated with the heat dissipated by the hot particle, and the heat absorbed by the cold particle such that the energy balance condition is automatically satisfied. Indeed it is possible to show, 
\begin{equation}
    \label{eq:ap_two_temp_Q_enp}
   \langle \Delta \dot{S}_{m(\tau)}^{tw} \rangle =  -\frac{\langle \dot{Q}_1 \rangle}{(T + \Delta T)} -  \frac{\langle \dot{Q}_2 \rangle}{T} =k_B \frac{k_2^2 \Delta T^2 \epsilon^2 (1-\epsilon^2)}{(k_1 + k_2) \gamma T(T + \Delta T) }.
  \end{equation}
Note that this expression agrees with that of $\Pi_{tw}$ (in units of $k_B/s$) in the main text (Eq.(7)).
It's an open and very interesting question if such an approach as we take here can be generalised to systems with more than two degrees of freedom with reservoirs at different temperatures, if the diffusion matrix is also non-diagonal. The expression for the medium entropy Eq.\ \eqref{eq:mediumep_path} always holds. But to identify the individual terms as heat flows requires the extra step that we take in Eq.\ \eqref{eq:SQrel_two}. This condition is insufficient to identify all individual heat flows,  if there are more than two degrees of freedom.

\subsection{ Details of numerical simulations}
\label{ap5:numerical}
The coupled \textit{Langevin} equations corresponding to the hydrodynamically coupled particles are linear first-order stochastic differential equations with appropriate additive noise terms(Eq.~\eqref{ap:eq_langevin}). Importantly, the diffusion matrices ($\mathbf{D}$) for the systems with hydrodynamic interactions are not diagonal, so the matrices incorporating the strength of the appropriate noise terms ($\mathbf{G}$) are calculated after performing the Cholesky decomposition such that $\mathbf{D} = \frac{1}{2} \mathbf{GG}^T$~\cite{thapa2024nonequilibrium}.  Then the differential equations (Eq.~\eqref{ap:eq_langevin}) with system-dependent drift and diffusion matrices are numerically discretized with fixed time step $\Delta t = 5\times10^{-5}s$ (which is less than all timescales present in the systems) and the integration is performed with the 1st-order Eurler-Maruyama method.  \\
For the \textit{OU-noise driven model}, the noise matrix $\mathbf{G}$ will be of the form $\mathbf{G} = \begin{pmatrix}
    \sqrt{2 D_0}& 0& 0\\
    \epsilon \sqrt{2 D_0} & \sqrt{2(D_0-\epsilon^2D_0)} & 0\\
    0& 0& \sqrt{2 D_e/\tau_e^2}   
\end{pmatrix} $. Then the dynamical equations (Eq.~\eqref{ap:eq_langevin}) can be explicitly written in the following discretized form,
\begin{align}
    \begin{split}
        x_1^{t+\Delta t} &= x_1^t - (k_1/\gamma)\ x_1^t \Delta t - (\epsilon k_2/\gamma)\ x_2^t \Delta t + (k_1/\gamma)\ \lambda^t \Delta t + \sqrt{2D_0 \Delta t}\ \eta_1^t, \\
        x_2^{t+\Delta t} &= x_2^t - (\epsilon k_1/\gamma)\ x_1^t\Delta t - (k_2/\gamma)\ x_2^t\Delta t + (\epsilon k_1/\gamma)\ \lambda^t\Delta t  + \epsilon \sqrt{2D_0 \Delta t}\ \eta_1^t + \sqrt{2(D_0-\epsilon^2D_0)\Delta t }\ \eta_2^t, \\
        \lambda^{t + \Delta t} &= \lambda^t -(1/\tau_e)\ \lambda^t \Delta t +  \sqrt{2 (D_e/\tau_e^2) \Delta t}\ \eta_3^t,
    \end{split}
    \label{ap:eq_ou_lan_discrete}
\end{align}
where $\eta_i^t \sim \mathcal{N}(0,1)$ is sampled from a Gaussian distribution with zero mean and unit variance and the initial states ($x_1^0, x_2^0, \lambda^0$) of the individual degrees of freedom are sampled from normal distributions with zero mean and appropriate variances such that, $x_1^0\sim \mathcal{N}(0,\frac{k_B T}{k_1})$,   $x_2^0\sim \mathcal{N}(0,\frac{k_B T}{k_2})$ and $\lambda^0\sim \mathcal{N}(0,\frac{D_e}{\tau_e})$. \\

Now, for the \textit{two-temperature model}, the drift ($\mathbf{F}_{tw}$) and diffusion matrix ($\mathbf{D}_{tw}$) corresponding to the \textit{langevin} dynamics are of the forms: $
    \mathbf{F}_{tw} = 
    \begin{pmatrix}
        k_1/\gamma & \epsilon k_2/\gamma\\
       \epsilon k_1/\gamma & k_2/\gamma  \\
       \end{pmatrix},\
   \mathbf{D}_{tw} = \begin{pmatrix}
      \frac{k_B (T + \Delta T)}{\gamma} &  \epsilon \frac{k_B (T + \Delta T)}{\gamma}\\
     \epsilon \frac{k_B (T + \Delta T)}{\gamma}  &   \frac{k_B (T +  \epsilon^2 \Delta T)}{\gamma}\\
       \end{pmatrix}
$. The noise matrix ($\mathbf{G}_{tw}$) can be computed as $\mathbf{G}_{tw} = \begin{pmatrix}
    \sqrt{2k_B(T+\Delta T)/\gamma} & 0\\
    \epsilon \sqrt{2k_B(T+\Delta T)/\gamma} & \sqrt{2 k_B T (1-\epsilon^2)/\gamma}
\end{pmatrix}$.  \\
The discrete-time evolution of the dynamics can be written as follows,
\begin{align}
    \begin{split}
        x_1^{t+\Delta t} &= x_1^t - (k_1/\gamma)\ x_1^t \Delta t - (\epsilon k_2/\gamma)\ x_2^t \Delta t + \sqrt{\Delta t \ 2k_B(T+\Delta T)/\gamma }\ \eta_1^t, \\
        x_2^{t+\Delta t} &= x_2^t - (\epsilon k_1/\gamma)\ x_1^t\Delta t - (k_2/\gamma)\ x_2^t\Delta t  + \epsilon \sqrt{\Delta t \ 2k_B(T+\Delta T)/\gamma }\ \eta_1^t + \sqrt{\Delta t\ 2 k_B T (1-\epsilon^2)/\gamma}\ \eta_2^t.
    \end{split}
    \label{eq:ap_two_lan_discrete}
\end{align}
Similar to the previous model, $\eta_i^t \sim \mathcal{N}(0,1)$ is sampled from a Gaussian distribution with zero mean and unit variance and the initial states ($x_1^0, x_2^0$) of the individual degrees of freedom are sampled from normal distributions with zero mean and appropriate variances such that, $x_1^0\sim \mathcal{N}(0,\frac{k_B T}{k_1})$,   $x_2^0\sim \mathcal{N}(0,\frac{k_B T}{k_2})$. 

\end{widetext}
\end{document}